\documentclass[%
 reprint,
superscriptaddress,
longbibliography,
amsmath,amssymb,
aps,
physrev
]{revtex4-2}

\usepackage{graphicx}
\usepackage{dcolumn}
\usepackage{bm}
\usepackage{xcolor}
\usepackage{comment}
\usepackage{slashed}

\usepackage[T1]{fontenc}
\usepackage{makecell}

\setcellgapes{4pt}
\usepackage{booktabs,multirow}

\usepackage{array}
\newcommand{\PreserveBackslash}[1]{\let\temp=\\#1\let\\=\temp}
\newcolumntype{C}[1]{>{\PreserveBackslash\centering}p{#1}}
\newcolumntype{R}[1]{>{\PreserveBackslash\raggedleft}p{#1}}
\newcolumntype{L}[1]{>{\PreserveBackslash\raggedright}p{#1}}

\graphicspath{{Figures/}}

\newcommand{\AVS}{$A$V$_3$Sb$_5$}

\newcommand{\CoSnS}{Co$_3$Sn$_2$S$_2$}

\newcommand{\RhInS}{Rh$_3$In$_2$S$_2$}
\newcommand{\RhInSe}{Rh$_3$In$_2$Se$_2$}
\newcommand{\RhSnS}{Rh$_3$Sn$_2$S$_2$}
\newcommand{\RhSnSe}{Rh$_3$Sn$_2$Se$_2$}
\newcommand{\RhTlS}{Rh$_3$Tl$_2$S$_2$}
\newcommand{\RhTlSe}{Rh$_3$Tl$_2$Se$_2$}
\newcommand{\RhPbS}{Rh$_3$Pb$_2$S$_2$}
\newcommand{\RhPbSe}{Rh$_3$Pb$_2$Se$_2$}
\newcommand{\RhBiS}{Rh$_3$Bi$_2$S$_2$}
\newcommand{\RhBiSe}{Rh$_3$Bi$_2$Se$_2$}

\newcommand{\PdInS}{Pd$_3$In$_2$S$_2$}
\newcommand{\PdInSe}{Pd$_3$In$_2$Se$_2$}
\newcommand{\PdSnS}{Pd$_3$Sn$_2$S$_2$}
\newcommand{\PdSnSe}{Pd$_3$Sn$_2$Se$_2$}
\newcommand{\PdTlS}{Pd$_3$Tl$_2$S$_2$}
\newcommand{\PdTlSe}{Pd$_3$Tl$_2$Se$_2$}
\newcommand{\PdPbS}{Pd$_3$Pb$_2$S$_2$}
\newcommand{\PdPbSe}{Pd$_3$Pb$_2$Se$_2$}
\newcommand{\PdBiS}{Pd$_3$Bi$_2$S$_2$}
\newcommand{\PdBiSe}{Pd$_3$Bi$_2$Se$_2$}

\newcommand{\NiInS}{Ni$_3$In$_2$S$_2$}

\usepackage{hyperref}
\usepackage{cleveref}

\begin{document}


\title{Electronic and structural properties of Rh- and Pd-based kagome-layered shandites from first principles}

\author{Luca Buiarelli}
\affiliation{Niels Bohr Institute, University of Copenhagen, 2200 Copenhagen, Denmark}
\affiliation{Department of Chemical Engineering and Materials Science, University of Minnesota, MN 55455, USA}

\author{Turan Birol}
\affiliation{Department of Chemical Engineering and Materials Science, University of Minnesota, MN 55455, USA}

\author{Brian M. Andersen}
\affiliation{Niels Bohr Institute, University of Copenhagen, 2200 Copenhagen, Denmark}

\author{Morten H. Christensen}
\email{mchriste@nbi.ku.dk}
\affiliation{Niels Bohr Institute, University of Copenhagen, 2200 Copenhagen, Denmark}

\date{\today}

\begin{abstract}
The shandite structure hosts transition metal ions arranged in kagome layers. These layers are stacked rhombohedrally and are interspersed with post-transition metal ions and chalcogens. The electronic states near the Fermi level are dominated by the transition metal $d$-orbitals and feature saddle points near several of the high-symmetry positions of the Brillouin zone, most notably the F and L points. Combining symmetry considerations with \emph{ab initio} methods, we study the electronic and structural properties of these materials with an emphasis on the connection between electronic saddle points at specific momenta and structural instabilities at these momenta. While the parent compounds studied are all found to be structurally stable under ambient conditions, we show that, in specific compounds, moving the saddle point closer to the Fermi level using either hydrostatic pressure or doping, can induce a structural instability. The importance of the electronic degrees of freedom in driving this instability is supported by the dependence of the frequency of the soft phonon mode on the electronic smearing temperature. Our first-principles calculations show that as the smearing temperature is increased, the compound becomes structurally stable again, indicating that the electron-phonon coupling is playing an important role. Our findings survey the structural properties of a large family of shandite materials and shed light on the role played by saddle points in the electronic structure in driving structural instabilities in rhombohedrally stacked kagome-layered materials.
\end{abstract}

\maketitle

\section{Introduction}\label{sec:intro}

Kagome-layered materials have attracted a great deal of attention in the condensed matter physics and materials science communities recently~\cite{Yin2022Topological,Neupert2022Charge}. This increased interest has been engendered, in part, by the discovery of a giant anomalous Hall effect in \CoSnS{}~\cite{Liu2018Giant}, superconductivity in the alkali-based \AVS{} kagome metals~\cite{Ortiz2020CsV3Sb5,Ortiz2021Superconductivity,Yin2021Superconductivity}, and exotic charge orders in ScV$_6$Sn$_6$~\cite{Arachchige2022Charge} and FeGe~\cite{Teng2022Discovery}. Such a wide range of materials provide a fertile ground for studying the interplay between electronic correlations and lattice geometry. In particular, many of these compounds exhibit temperature-driven structural transitions accompanied by some type of charge order~\cite{Arachchige2022Charge,Ortiz2019New,Jiang2021Unconventional,Shumiya2021Intrinsic,Mielke2022Time-reversal,Stahl2022Temperature-driven,Teng2022Discovery,Teng2023Magnetism}. This raises the question of the relative role of the phononic and electronic degrees of freedom in driving these transitions: The electronic structure of the kagome lattice features van Hove points at multiple high-symmetry points of the Brillouin zone which promote the effects of correlations~\cite{vanHove1953Occurrence,Kiesel2013Unconventional}. In spite of this, the different materials exhibit distinct behaviors and the role played by electrons in driving the observed instabilities remains an open question.

In the \AVS{} materials, angle-resolved photo-emission spectroscopy (ARPES) measurements and density-functional theory (DFT) are in overall agreement~\cite{Ortiz2020CsV3Sb5,Kang2022Twofold,Hu2022Rich,Li2022Coexistence}. Both show an electronic structure with multiple van Hove points near the M point of the Brillouin zone~\cite{Tan2021Charge,Kang2022Twofold}. These are of both `pure' and `mixed' types, denoting the localization of the electronic wavefunction within one unit cell; at pure van Hove points, the wavefunction is localized on a single atom of the kagome lattice, while at mixed van Hove points, it is localized on two of the kagome atoms~\cite{Kang2022Twofold,Kiesel2012Sublattice}. The pure versus mixed nature of the Bloch wavefunction has important consequences for the leading instabilities and the nature of the resulting ordered state~\cite{Yu2012,Kiesel2013Unconventional,Wang2013,Wu2021Nature,Lin2022,Tazai,Romer2022,Wu2023Sublattice,Holbaek2023Unconventional,YiDai2024,Li2024Intertwined}. Charge order onsets near 100~K in all three \AVS{} compounds with a concomitant breaking of the translational symmetry and enlargement of the unit cell along the $a$-, $b$-, and $c$-directions~\cite{Stahl2022Temperature-driven,Wu2022Charge,Kautzsch2023Structural,Scagnoli2024Resonant}. There are indications that the charge order breaks time-reversal symmetry~\cite{Jiang2021Unconventional,Mielke2022Time-reversal,Khasanov2022Time-reversal,Xu2022Three-state,Guo2022Switchable,Xing2024Optical,ChristensenLoop}, although these observations remain controversial due to conflicting reports~\cite{Li2022Rotation,Saykin2023High,Farhang2023Unconventional}. It is tempting to associate the charge order with an electronic instability driven by nesting of the large densities of states at the van Hove points~\cite{Park2021Electronic,Lin2021Complex,Christensen2021Theory}. However, the three-dimensional nature of the materials implies that van Hove points are associated with kinks in the density of states -- rather than true divergences as in the two-dimensional case~\cite{vanHove1953Occurrence} -- matching observations of at most a moderate increase in the density of states at the van Hove points~\cite{Zhao2021Cascade}. Instead, the instability appears to be a consequence of the interplay between electronic and phononic degrees of freedom and it is absent if the two systems are decoupled~\cite{Tan2021Charge,Ratcliff2021Coherent,Christensen2021Theory,Subedi2022Hexagonal,Gutierrez-Amigo2024Phonon}. Indeed, even the anharmonic couplings between phonon modes turn out to be important in stabilizing the charge density wave phase~\cite{Ritz2023Impact}. These results are consistent with how charge density waves emerge in most metals~\cite{Johannes2008Fermi}. Charge order in FeGe emerges deep within an antiferromagnetic phase~\cite{Teng2022Discovery}. Similar to \AVS{}, ARPES finds multiple van Hove points near the Fermi level~\cite{Teng2023Magnetism}, and the ordering vectors of the charge order are the same in both FeGe and \AVS{}~\cite{Teng2022Discovery,Yin2022Discovery}. However, the ARPES results are not faithfully reproduced by DFT calculations~\cite{Teng2023Magnetism}, indicating that correlations in FeGe play a prominent role in shaping the electronic structure, and might also drive the formation of a charge-ordered phase~\cite{Wu2024Symmetry,Bonetti2024Competing}. The role of the van Hove points in shaping the physics of both \AVS{} and FeGe is in stark contrast to the phenomenology observed in ScV$_6$Sn$_6$~\cite{Arachchige2022Charge}. In this case, the charge order leads to a $\sqrt{3} \times \sqrt{3}$ enlargement of the unit cell driven by lattice fluctuations~\cite{Lee2024Nature,Korshunov2023Softening}. This dichotomy raises the question of the role played by saddle points in the electronic band structure in driving structural transitions in kagome-layered materials.

Shandite materials offer a fertile ground for such a study. These compounds feature kagome layers of transition metals and their electronic structures near the Fermi level are dominated by the $d$-electrons of these transition metals. It is a very diverse family with the generic chemical composition $M_3 A_2$Ch$_2$, where $M$ is a transition metal (Co, Ni, Rh, or Pd), $A$ is a post-transition metal (In, Sn, Tl, Pb, or Bi) while Ch denotes one of the chalcogenides, S or Se. Together with cubic and monoclinic parkerites~\cite{Range1983Neue}, shandites are part of the larger family of half-antiperovskites~\cite{Weihrich2006Half}. Although the monoclinic parkerites are quasi-layered, they do not feature kagome layers and they will not be the focus of this work. The most famous shandite is the ferromagnetic \CoSnS{}~\cite{Schnelle2013Ferromagnetic} which features a host of exotic phenomena from giant anomalous Hall~\cite{Liu2018Giant,Wang2018Large} and Nernst~\cite{Yang2020Giant} effects to emergent Weyl fermions~\cite{Morali2019Fermi-arc,Liu2019Magnetic} and enhanced magneto-optical responses~\cite{Okamura2020Giant}. Magnetism can be continuously suppressed by substituting Sn for In~\cite{Pielnhofer2014Half,Corps2015Interplay} or Co by Ni~\cite{Weihrich2004Magnetischer} and magnetic shandites are, to the best of our knowledge, still rare. More recently, the nonmagnetic compounds \NiInS{} and Ni$_3$In$_2$Se$_2$ have been shown to feature a large magnetoresistance induced by Dirac nodal lines in the electronic structure~\cite{Zhang2022Endless, Cao2023Crystal, Kumar2024Endless, Zhai2025Highly}. 

We focus on the less studied shandite variants based on Rh and Pd. These may be harder to synthesize in an $R\bar{3}m$ crystal structure compared to their Ni- and Co-based counterparts. Ref.~\cite{Zabel1979Ternary} reports the successful growth of five such compounds in the shandite structure, Pd$_3$Tl$_2$S$_2$, Pd$_3$Tl$_2$Se$_2$, Pd$_3$Pb$_2$Se$_2$, Rh$_3$Tl$_2$S$_2$, and Rh$_3$Pb$_2$S$_2$, in addition to two Co-based compounds and eight Ni-based ones. Subsequently, Rh$_3$In$_2$S$_2$, Rh$_3$Pb$_2$Se$_2$, and Rh$_3$(Pb,In)$_2$S$_2$ were also grown in the shandite structure~\cite{Natarajan1988Synthesis}. More recent studies have reproduced the shandite structure of Pd$_3$Pb$_2$Se$_2$~\cite{Seidlmayer2010Half,Yu2020Pressure-induced}. Interestingly, Ref.~\cite{Yu2020Pressure-induced} conducted a pressure-study of this compound up to nearly 80 GPa under which the crystal maintains the $R\bar{3}m$ structure, but becomes superconducting near 26.5 GPa. The only report of Pd$_3$Pb$_2$S$_2$ is as a naturally occurring mineral, laflammeite, with a monoclinic $C2/m$ parkerite structure~\cite{Barkov2002Laflammeite}, although previous first-principles results find a shandite structure~\cite{Nie2022Tunable,Basak2025Dynamical}. It is also worth noting that the rhombohedral $R\bar{3}m$ of Co$_3$In$_2$S$_2$ has previously been mistaken as monoclinic~\cite{Michelet1976The,Clauss1978Crystal-chemical}. On the other hand, for compounds containing Bi, only the parkerite structure has been observed so far~\cite{Range1983Neue,Baranov2001Crystal,Sakamoto2007Charge-density-wave,Sakamoto2007Transport,Kaluarachchi2015Superconductivity,Lapano2021,Zhang2024Structure}, with most of them falling in the $C2/m$ space group, although Pd$_3$Bi$_2$S$_2$ is observed in the cubic $I2_1 3$ group~\cite{Weihrich2007Structure,Seidlmayer2010Half}. Interestingly, among the Pd- and Rh-based parkerites, superconductivity is observed in both Pd$_3$Bi$_2$Se$_2$~\cite{Sakamoto2007Transport,Weihrich2011Palladium,Lapano2021} and Rh$_3$Bi$_2$Se$_2$~\cite{Sakamoto2007Charge-density-wave}. Many of the compounds we consider in this study are thus reported to exist in the shandite crystal structure. However, in order to evaluate trends in this structure family, we also consider compounds that are either not yet synthesized, or reported in one of the two parkerite structures. As detailed in Sec.~\ref{sec:first_principles}, both parkerite and shandite structures could be metastable for most compounds studied. Importantly, neither of the two parkerite structures can be obtained from the shandite by continuous crystal structural changes that can be explained by dynamically unstable phonons. In other words, while either shandite or parkerite structures may be (meta)stable, a transition between the two is not expected.

In this work, we map out the electronic structures of a large family of shandite compounds with the goal of understanding the role of van Hove points in driving possible structural instabilities of these materials. In the Peierls scenario, a structural transition is a secondary effect of a charge-density wave (CDW) transition driven by electronic nesting, which takes place independent of the lattice. In real materials however, the electronic charge order and the structural transition are heavily intertwined and the two transitions occur simultaneously. Consequently, electronic nesting alone is insufficient for driving a CDW transition in most real materials~\cite{Johannes2008Fermi}, in contrast to the Peierls scenario. In particular, a sufficiently strong electron-phonon coupling is required~\cite{Yeats1976Peierls}. This begs the question of the relative role played by electronic and structural degrees of freedom in driving CDW transitions in kagome materials. The saddle points present in the electronic structure of kagome-layered materials enhance the effects of electronic correlations and phenomena driven by electronic nesting~\cite{vanHove1953Occurrence,Nandkishore2012Chiral,Kiesel2012Sublattice,Kiesel2013Unconventional,Classen2020Competing}. However, as discussed above, this does not seem to be the full picture in the \AVS{} kagome metals, where the interplay between electrons and phonons is crucial for capturing the CDW transition~\cite{Tan2021Charge,Ratcliff2021Coherent,Christensen2021Theory,Subedi2022Hexagonal,Gutierrez-Amigo2024Phonon,Ritz2023Impact}. Here, we examine how electronic nesting impacts the possible structural instabilities in shandite materials. In this context, a recent first-principles study of the dynamic properties of (Pb,Pt)$_3$Pb$_2$Ch$_2$ found a transition from the shandite $R\bar{3}m$ phase to an $R\bar{3}c$ phase in Pt$_3$Pb$_2$S$_2$ while the remaining three compounds were found to be structurally stable~\cite{Basak2025Dynamical}. We extend this framework and consider 20 possible variations of the shandites and calculate their electronic and phononic spectra. Tuning the chemical potential through pressure or simulated doping, we attempt to induce structural transitions and establish how these are impacted by the electrons.

The paper is organized as follows: In Sec.~\ref{sec:crystal_structure} we review the geometry of the shandite crystal structure and in Sec.~\ref{sec:phenomenology} we present a phenomenological study of the structural instabilities possible in the shandites. The results of our first-principles calculations are presented in Sec.~\ref{sec:first_principles} which also includes a study of the symmetry properties of the electronic bands. In Secs.~\ref{sec:stable_compounds} and \ref{sec:unstable_compounds}, we present detailed \emph{ab initio} calculations of the effects of pressure and doping on specific shandite compounds. In Sec.~\ref{sec:conclusions}, we present our conclusions and discuss possible directions for future research.

\section{Shandite Crystal structure}\label{sec:crystal_structure}

\begin{figure}[t]
    \centering
    \includegraphics[width=\columnwidth]{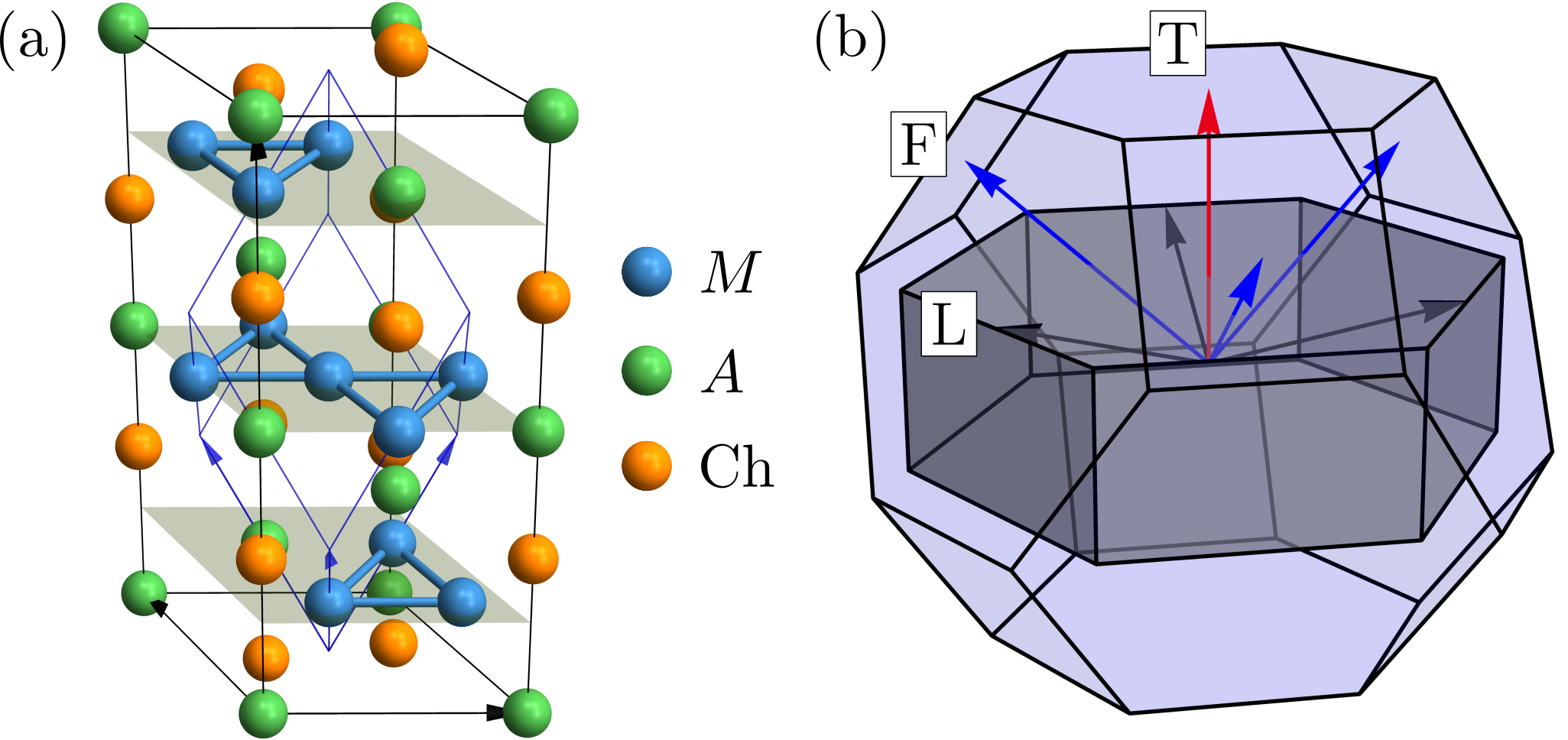}
    \caption{\label{fig:r3m_crystal_and_bz} (a) Schematic crystal structure of the shandite compounds. The rhombohedral primitive unit cell is denoted by the light blue arrows, while the hexagonal conventional cell is outlined in black. The kagome layers consist of the transition-metal ions, $M$, in deep blue. The post transition-metal ions, $A$, are green while the chalcogen ions, Ch, are orange. (b) First Brillouin zone for the space group $R\bar{3}m$. The Brillouin zone of the primitive unit cell is denoted in light blue, while the dark gray denotes the Brillouin zone of the conventional cell. The relevant high-symmetry points are highlighted by blue (F), black (L), and red (T) arrows, respectively. The stars of F and L contain three points while there is only a single point in the star of T.}
\end{figure}

Shandites are rhombohedral materials that crystallize in the $R\bar{3}m$ space group (\#166) with point group D$_{\rm 3d}$. This group permits perfect (i.e., non-distorted) kagome layers due to the presence of the three-fold roto-inversion symmetry $\bar{3}$, which is equivalent to a $60^{\circ}$ rotation followed by a mirror in the plane perpendicular to the rotation axis. Such a kagome-layered structure is familiar from other materials, such as the quantum spin liquid candidate material herbertsmithite~\cite{Norman2016Colloquium} or the frustrated magnet Na$_2$Ti$_3$Cl$_8$ \cite{Haenni2017, Paul2020}. The crystal structure contains three symmetry-equivalent kagome layers in the conventional hexagonal unit cell stacked in an `ABCABC$\ldots$' fashion. This is in contrast to simpler cases such as the alkali-based kagome metals \AVS{} in space group $P6/mmm$ and point group D$_{\rm 6h}$, in which a proper six-fold symmetry is present and the unit cell contains a single kagome layer (so the layers are stacked in `AAA$\ldots$' fashion)~\cite{Ortiz2019New,Wu2022Charge,Kautzsch2023Structural,Ritz2023Impact}. As a result, the crystal structure of the shandite materials is more involved than that of the corresponding hexagonal $P6/mmm$ kagome compounds, as shown in Fig.~\ref{fig:r3m_crystal_and_bz}(a). However, a rhombohedral \textit{primitive} cell can be constructed that only contains one kagome layer, outlined by the blue cell in Fig.~\ref{fig:r3m_crystal_and_bz}(a). In terms of the conventional lattice parameters, the primitive translations are
\begin{align}
\label{eq:transl_rh}
    \mathbf{R}_1 &= \tfrac{1}{3}(  0,  \sqrt{3}a,  c)\,, \nonumber \\
    \mathbf{R}_2 &= \tfrac{1}{6}( 3a, -\sqrt{3}a, 2c)\,, \\
    \mathbf{R}_3 &= \tfrac{1}{6}(-3a,  \sqrt{3}a, 2c)\,. \nonumber
\end{align}

An example of a shandite crystal structure is shown in Fig.~\ref{fig:r3m_crystal_and_bz}(a) with the kagome layers defined by the blue ions. The primitive rhombohedral unit cell is outlined by the light blue arrows while the black arrows denote the conventional hexagonal unit cell. The corresponding Brillouin zones are shown superimposed in Fig.~\ref{fig:r3m_crystal_and_bz}(b) and the high-symmetry points F, L, and T that will be relevant in this work, are indicated by arrows. In terms of the primitive reciprocal lattice vectors,
\begin{align}
    \mathbf{G}_1 &= 2\pi( \tfrac{1}{a},  \tfrac{\sqrt{3}}{a},  0)\,, \nonumber\\
    \mathbf{G}_2 &=  \pi( \tfrac{1}{a}, -\tfrac{\sqrt{3}}{a}, \tfrac{3}{c})\,, \\
    \mathbf{G}_3 &=  \pi(-\tfrac{3}{a}, -\tfrac{\sqrt{3}}{a}, \tfrac{3}{c})\,. \nonumber
\end{align}
these points are
\begin{align}
    {\rm F} &= (0,\tfrac{1}{2},\tfrac{1}{2})\,, \nonumber\\
    {\rm L} &= (0,\tfrac{1}{2},0)\,, \\
    {\rm T} &= (\tfrac{1}{2},\tfrac{1}{2},\tfrac{1}{2})\,. \nonumber
\end{align}
Applying the point group operations to these generates the star of each $k$-point, namely the set of distinct symmetry-equıvalent points in the Brillouin zone that are not reachable by a reciprocal lattice vector. These are summarized in Table~\ref{tab:BZ_summary}.
\begin{table}[t]
    \centering
    \begin{tabular}{C{0.1\linewidth}C{0.18\linewidth}C{0.28\linewidth}C{0.25\linewidth}C{0.13\linewidth}}
    \toprule
         BZ point & {\bf k}-vector (primitive) & Star of {\bf k} (primitive) & {\bf k}-vector (conventional) & Little group \\
         \hline
         \hline
         F & $(0,\tfrac{1}{2},\tfrac{1}{2})$ & \makecell{$\mathbf{Q}_{\rm F_1} = (0,\tfrac{1}{2},\tfrac{1}{2})$ \\[5pt] $\mathbf{Q}_{\rm F_2} = (\tfrac{1}{2},0,\tfrac{1}{2})$ \\[5pt] $\mathbf{Q}_{\rm F_3} = (\tfrac{1}{2},\tfrac{1}{2},0)$} & $(-\tfrac{1}{2},0,1)$ & \makecell{C$_{\rm 2h}$ \\ (2/m)} \\
         \midrule
         L & $(\tfrac{1}{2},0,0)$ & \makecell{$\mathbf{Q}_{\rm L_1} = (\tfrac{1}{2},0,0)$ \\[5pt] $\mathbf{Q}_{\rm L_2} = (0,\tfrac{1}{2},0)$ \\[5pt] $\mathbf{Q}_{\rm L_3} = (0,0,\tfrac{1}{2})$} & $(\tfrac{1}{2},\tfrac{1}{2},\tfrac{1}{2})$ &  \makecell{C$_{\rm 2h}$ \\ (2/m)} \\
         \midrule
         T & $(\tfrac{1}{2},\tfrac{1}{2},\tfrac{1}{2})$ & \makecell{$\mathbf{Q}_{\rm T_1}=(\tfrac{1}{2},\tfrac{1}{2},\tfrac{1}{2})$} & $(0,0,\tfrac{3}{2})$ & \makecell{D$_{\rm 3d}$ \\ ($\bar{3}$m)} \\
         \bottomrule
    \end{tabular}
    \caption{\label{tab:BZ_summary}Relevant high-symmetry points in the Brillouin zone (BZ). The first column shows the high-symmetry point, denoted by the colored arrows in Fig.~\ref{fig:r3m_crystal_and_bz}, and the second column shows their coordinate in the primitive basis. The third column denotes the star of each high-symmetry point. The {\bf k}-vector of the different points in the conventional unit cell is given in the fourth column, to be compared with the conventional Brillouin zone in Fig.~\ref{fig:r3m_crystal_and_bz}. In the last column, we include the little group of each of the high-symmetry points.}
\end{table}
From the table, it is clear that the vectors of the stars obey
\begin{align}
    \mathbf{Q}_{\mathrm{F}_i} + \mathbf{Q}_{\mathrm{F}_j} &= \epsilon_{ijk}\mathbf{Q}_{\mathrm{F}_k} \nonumber\\
    \mathbf{Q}_{\mathrm{L}_i} + \mathbf{Q}_{\mathrm{L}_j} &= \epsilon_{ijk}\mathbf{Q}_{\mathrm{F}_k} \label{eq:nesting}\\
    \mathbf{Q}_{\mathrm{F}_i} + \mathbf{Q}_{\mathrm{L}_j} &= \epsilon_{ijk}\mathbf{Q}_{\mathrm{L}_k}\,, \nonumber
\end{align}
modulo a reciprocal lattice vector and with $\epsilon_{ijk}$ the totally anti-symmetric Levi-Civita symbol.

In the shandites, the kagome layers are formed by transition-metal ions at the $9d$ Wyckoff positions (or $9e$ for a different origin choice) and the states near the Fermi level are dominated by the transition-metal $d$-electrons (see Sec.~\ref{sec:first_principles}). We focus on particular shandite materials which exhibit saddle points in the electronic dispersion near the F and/or L points shown in Fig.~\ref{fig:r3m_crystal_and_bz}(b). These saddle points are accompanied by van Hove peaks in the electronic density of states, although the three-dimensional nature of the shandites prevents the occurrence of true van Hove singularities. Consequently, due to the geometry of the Brillouin zone where the F and L points are connected by vectors in the star of either F or L [Eq.~\eqref{eq:nesting}], the regions with large densities of states are nested with each other. Hence, the shandites are promising candidate materials for examining the interplay between structural and electronic degrees of freedom in driving density-wave instabilities with wavevectors at F or L.

\section{Structural instabilities in $R\bar{3}m$ compounds}\label{sec:phenomenology}

\begin{figure}[!t]
    \centering
    \begin{minipage}[t]{0.21\textwidth}
    \vspace*{0pt}
        \centering
        \begin{tabular}{c|cc}
            $\mathbf{Q}_{\rm F_1}$ or $\mathbf{Q}_{\rm L_1}$ & $2_{100}$ & $\bar{1}$ \\
            \hline
             $F_{1}^{\pm}$ or $L_{1}^{\pm}$ & $+1$ & $\pm 1$ \\
             $F_{2}^{\pm}$ or $L_{2}^{\pm}$ & $-1$ & $\pm 1$ \\
        \end{tabular}
        
        \vspace*{5pt}
        \begin{tabular}{c|cc}
            $\mathbf{Q}_{\rm F_2}$ or $\mathbf{Q}_{\rm L_2}$  & $2_{010}$ & $\bar{1}$ \\
            \hline
             $F_{1}^{\pm}$ or $L_{1}^{\pm}$ & $+1$ & $\pm 1$ \\
             $F_{2}^{\pm}$ or $L_{2}^{\pm}$ & $-1$ & $\pm 1$ \\
        \end{tabular}
        
        \vspace*{5pt}
        \begin{tabular}{c|cc}
            $\mathbf{Q}_{\rm F_3}$ or $\mathbf{Q}_{\rm L_3}$  & $2_{110}$ & $\bar{1}$ \\
            \hline
             $F_{1}^{\pm}$ or $L_{1}^{\pm}$ & $+1$ & $\pm 1$ \\
             $F_{2}^{\pm}$ or $L_{2}^{\pm}$ & $-1$ & $\pm 1$ \\
        \end{tabular}
    \end{minipage}\begin{minipage}[t]{0.25\textwidth}
    \vspace*{0pt}
        \centering
        \includegraphics[width=\textwidth]{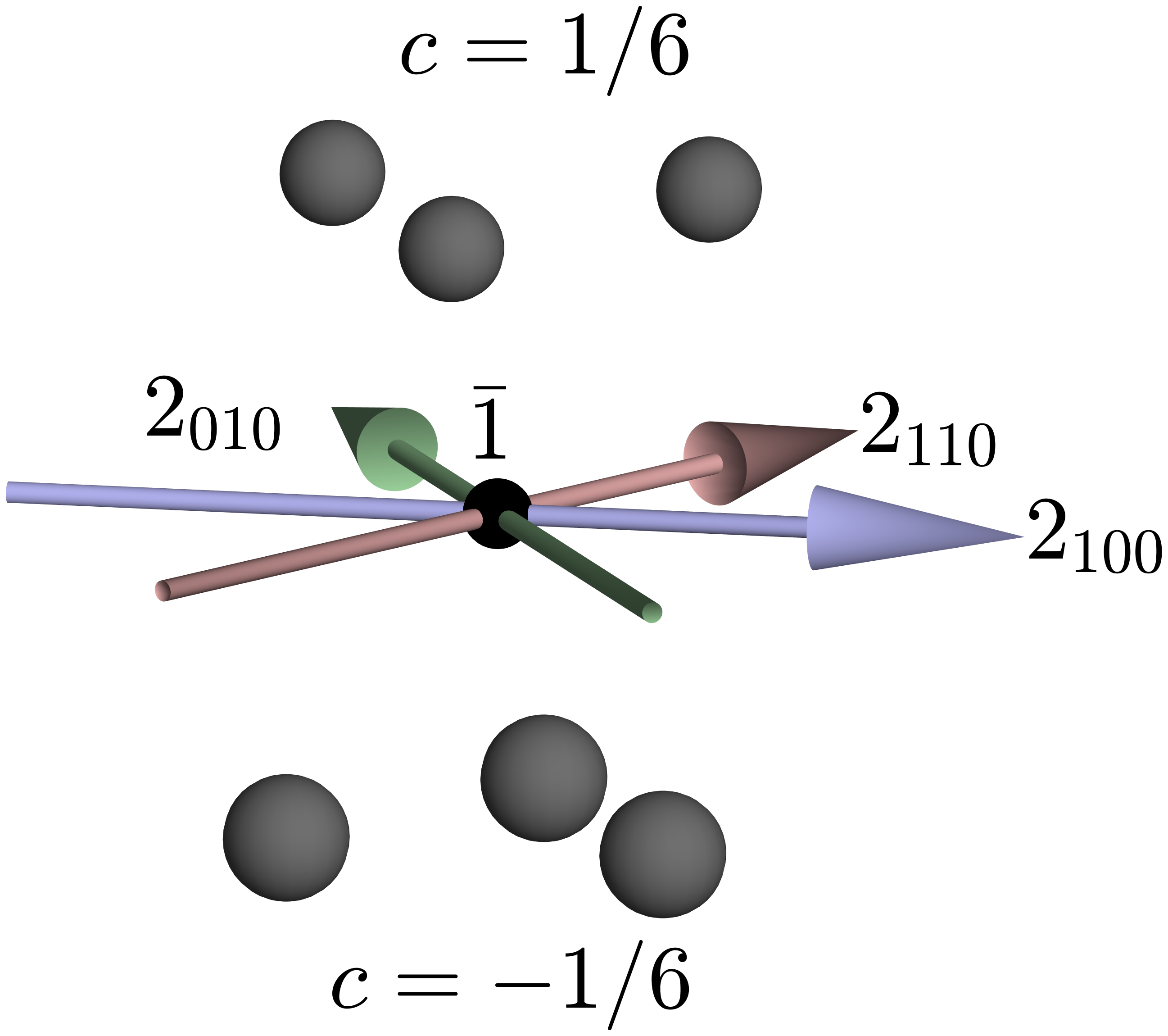}
    \end{minipage}
    \caption{\label{fig:r3barm_irreps}Characters of the irreps of the little group, C$_{\rm 2h}$, of the F or the L point of $R\bar{3}m$, alongside an illustration of the rotation axes and inversion point with respect to the kagome planes right above the origin at $c=1/6$ and right below the origin at $c=-1/6$. The table shows the characters of the generators. The little group irreps are one-dimensional, while the corresponding space group irreps are three-dimensional since there are three vectors in the star of F or L. The $\pm$ sign indicates the character under inversion, where the inversion center is placed at the origin of the conventional cell [Fig.~\ref{fig:r3m_crystal_and_bz}(a)] in between two kagome layers.}
\end{figure}

\begin{table*}[t]
    \centering
    \begin{tabular}{c||cccccc}
    \toprule
     Order & \makecell{1\textbf{Q}($F_1^+$) \\ (a;0;0)} & \makecell{3\textbf{Q}($F_1^+$) \\ (a;a;a)} & \makecell{1\textbf{Q}($L_2^-$) \\ (b;0;0)} & \makecell{3\textbf{Q}($L_2^-$) \\ (b;b;b)} & \makecell{1\textbf{Q}($F_1^+$)+2\textbf{Q}($L_2^-$) \\ (a;0;0|b;0;b)} & \makecell{3\textbf{Q}($F_1^+$)+3\textbf{Q}($L_2^-$) \\ (a;a;a|b;b;b)} \\
     \hline
     \makecell{Space \\ group} & $P2/m$ (\#10) & $R\bar{3}m$ (\#166) & $C2/m$ (\#12) & $R\bar{3}m$ (\#166) & $C2/m$ (\#12) & $R\bar{3}m$ (\#166) \\
     \hline
     \makecell{Unit \\ cell} & $1\times2\times1$ & $2\times2\times1$ & $1\times2\times2$ & $2\times2\times2$ & $2\times2\times2$ & $2\times2\times2$ \\
     \hline
     Ex. & 
     \makecell{\includegraphics[width=.145\textwidth]{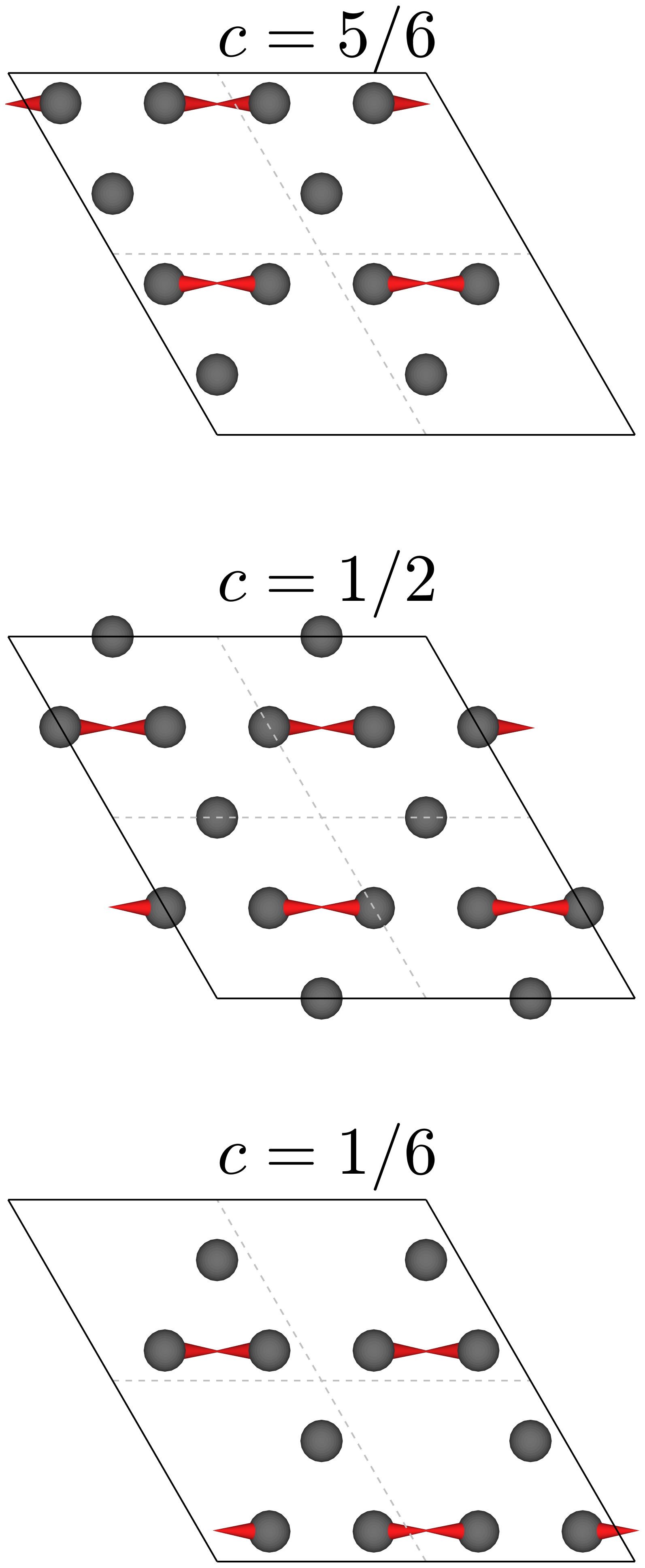}} & 
     \makecell{\includegraphics[width=.145\textwidth]{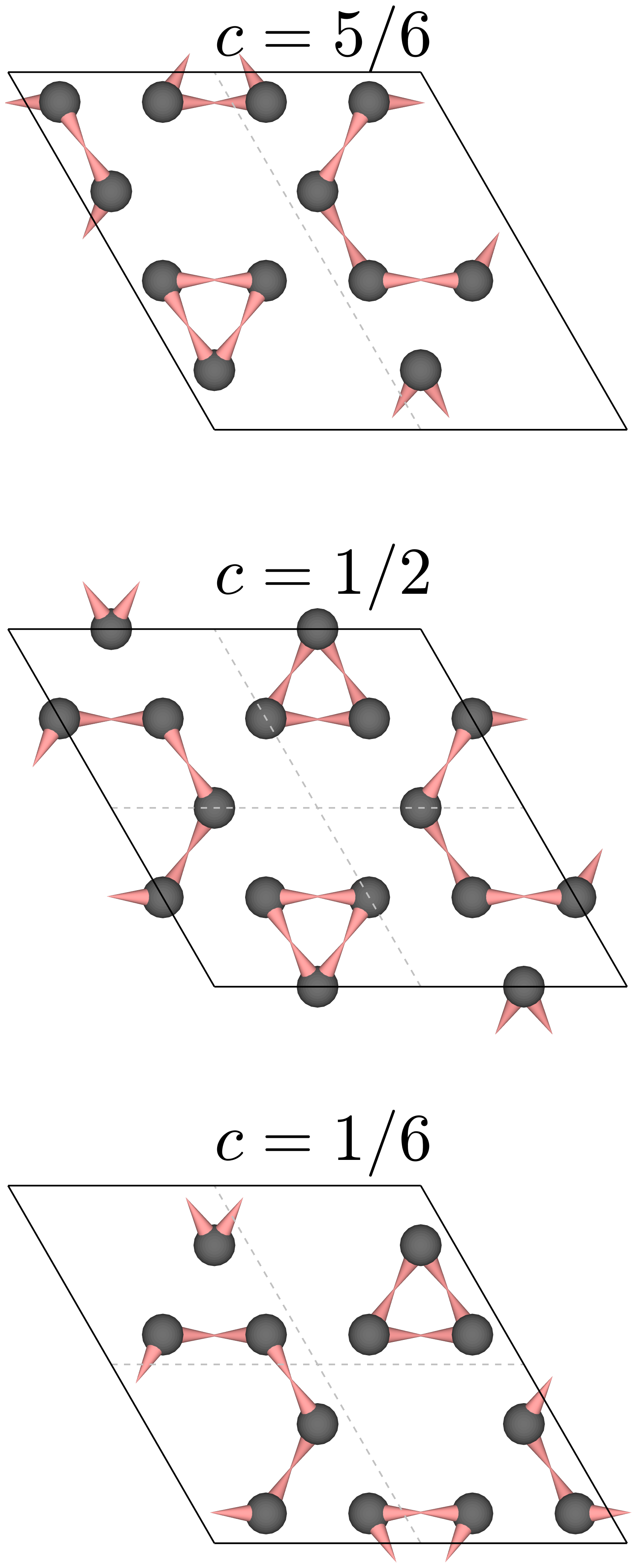}} & 
     \makecell{\includegraphics[width=.145\textwidth]{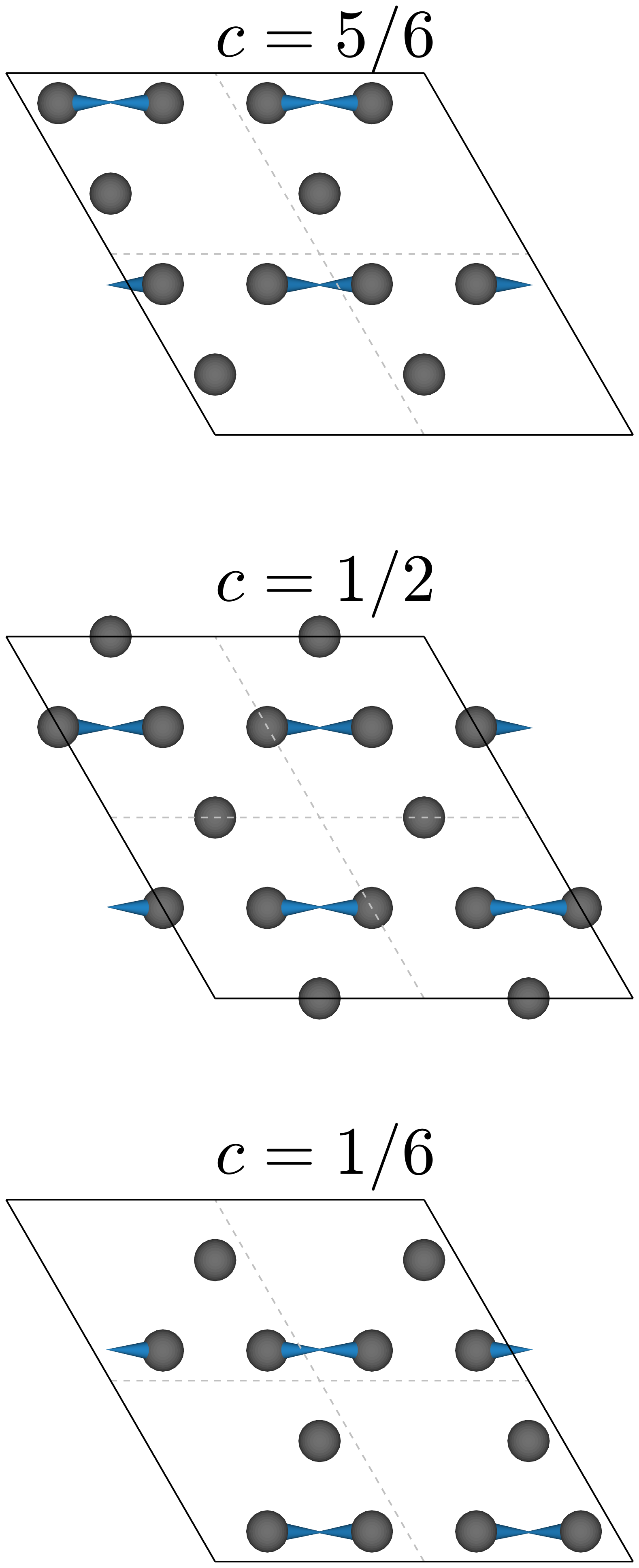}} &
     \makecell{\includegraphics[width=.145\textwidth]{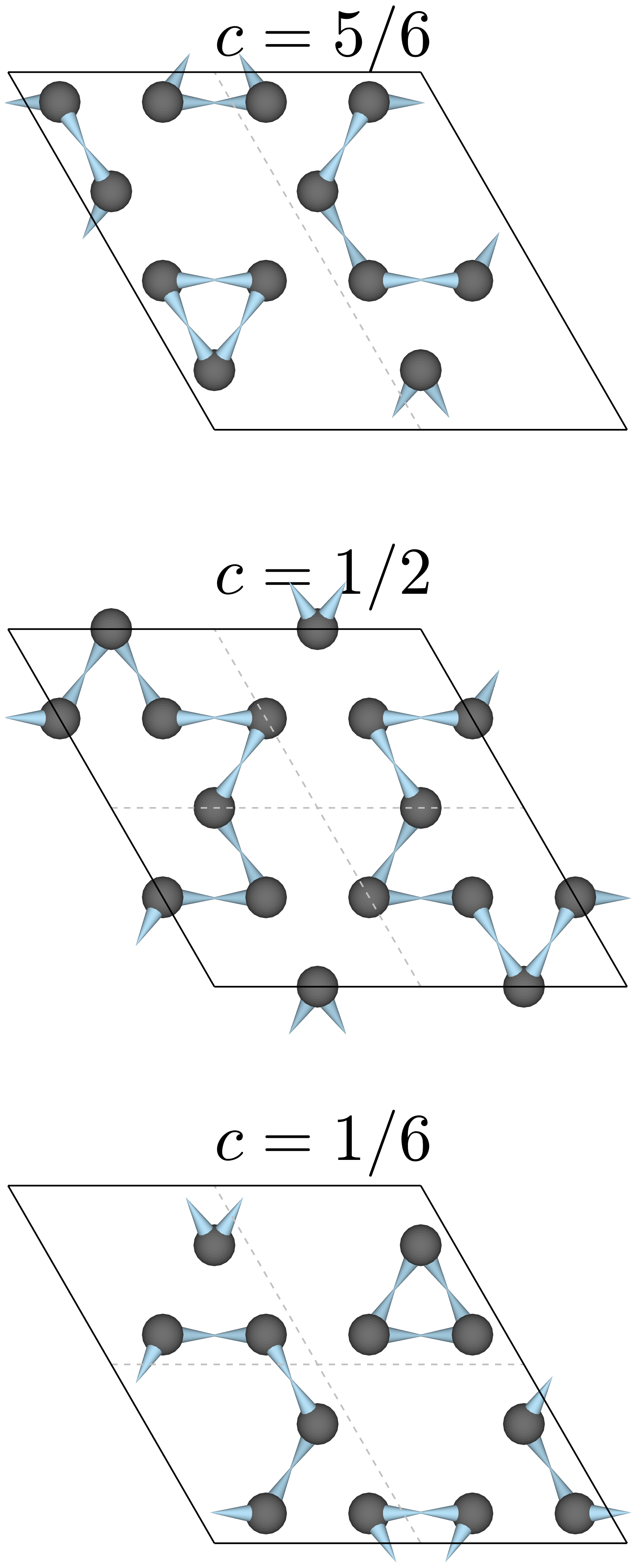}} &
     \makecell{\includegraphics[width=.145\textwidth]{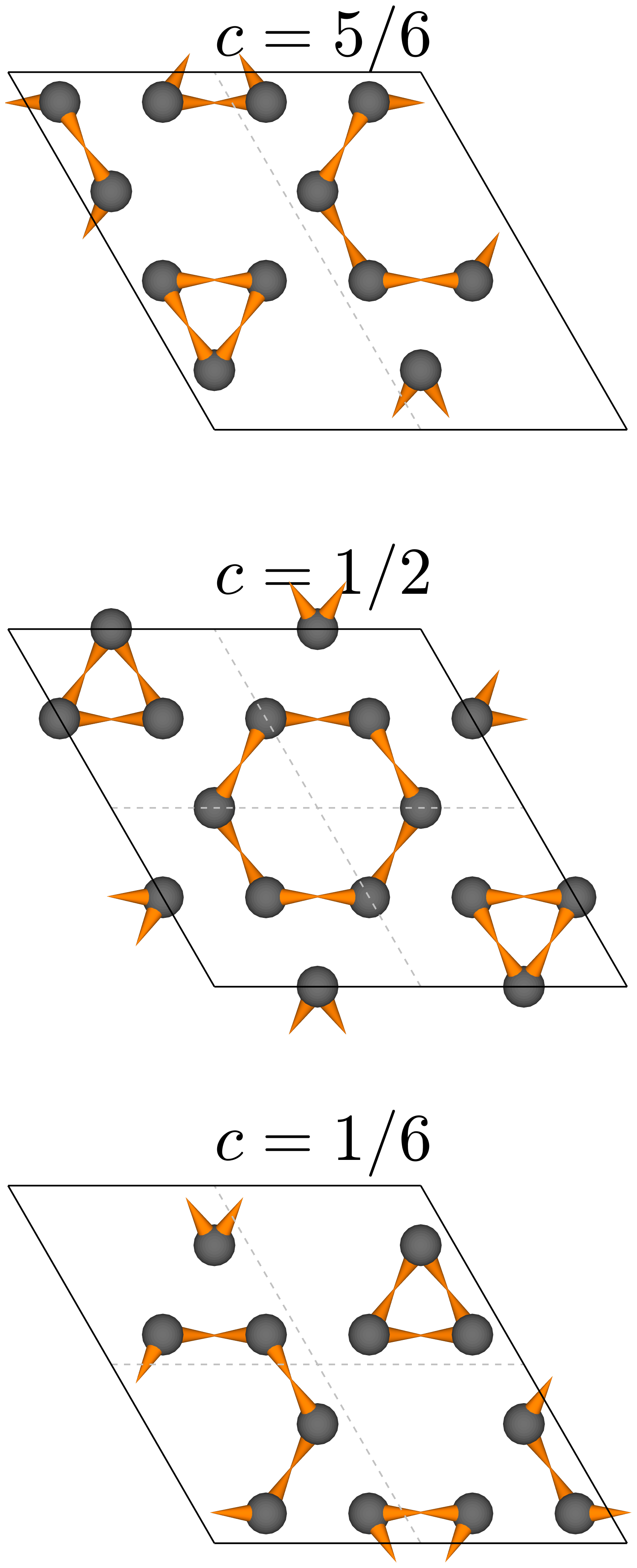}} &
     \makecell{\includegraphics[width=.145\textwidth]{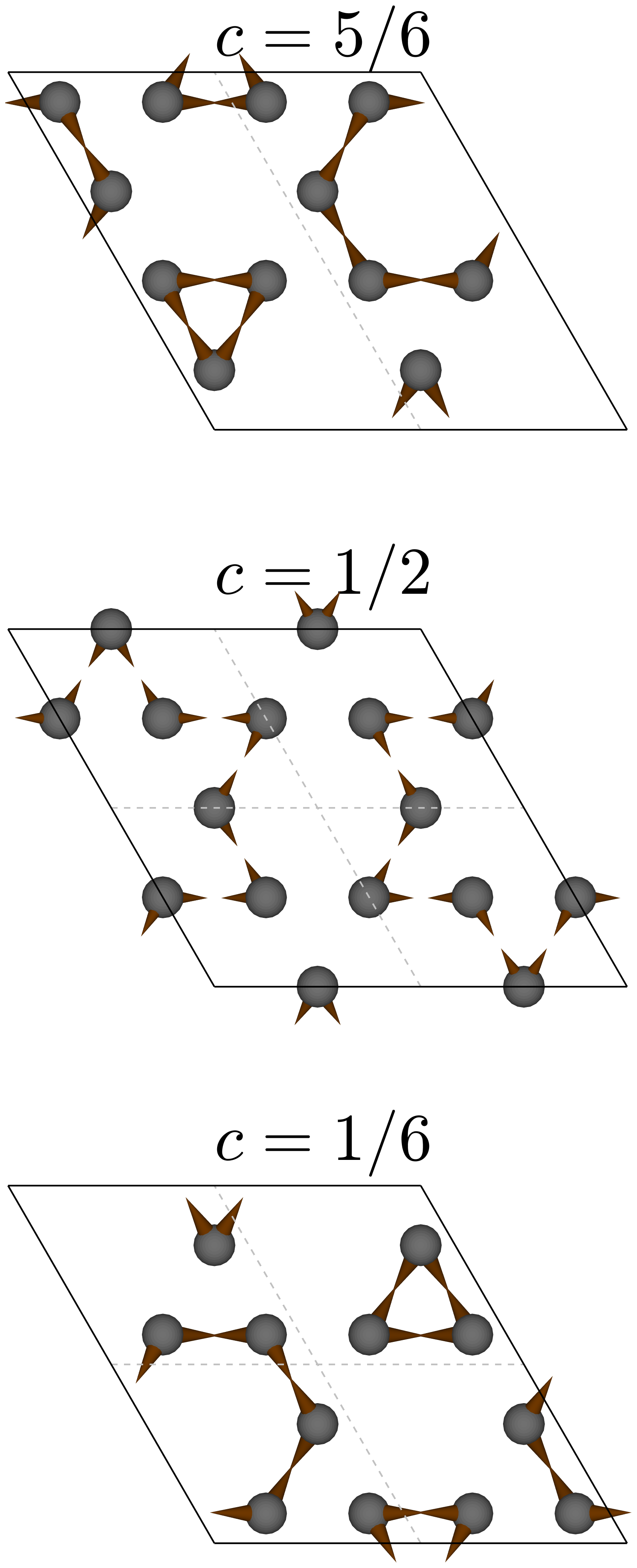}} \\
     \bottomrule
    \end{tabular}
    \caption{\label{tab:subgroups} Space groups resulting from the condensation of different orders that are candidate ground states of the free energy $\mathcal{F}_{F}+\mathcal{F}_L+\mathcal{F}_{FL}$ relevant for case (i) discussed in the text. The order parameters are shown as phonon displacements that dimerize two of the three kagome atoms. The three consecutive layers in the conventional unit cell are shown, at $c=1/6$, $c=1/2$ and $c=5/6$ along the conventional cell $c$ axis. Notice that the kagome planes are shifted by one of the vectors in Eq.~\eqref{eq:transl_rh} in consecutive layers. In the presence of $L$ order parameters, the unit cell doubles along the conventional $c$ axis, which results in a conventional cell containing six kagome layers. For simplicity, we illustrate only three layers; the remaining three can be envisioned as exhibiting the same phonon modes but with opposite sign.
}
\end{table*}

Here, we will focus on the phenomenology of instabilities occurring at the F or L points in the Brillouin zone due to the enhanced nesting between these points. The phenomenology of the transitions that we will consider can be described within Landau theory. The form of the Landau free energy depends solely on the symmetry properties of the order parameters, as given by their space group irreducible representations (irreps). In general, irreps of a symmetry group provide a complete list of different possible order parameters, and the space group irreps of the order parameters contain information on how, in addition to the breaking of the translational symmetry and the related enlargement of the primitive unit cell, the order parameters transform in different ways under the mirrors or rotations of the so-called little group. Consequently, the space group irreps for a specific $\mathbf{k}_{\ast}$-point are constructed from the little group and the star of $\mathbf{k}_{\ast}$. Here, the little group denotes the subgroup of the space group which leaves the wave vector at $\mathbf{k}_{\ast}$ invariant \cite{Bradley2010}. The little group of both the F and L points is C$_{\rm 2h}$; its generators and their characters are summarized in Fig.~\ref{fig:r3barm_irreps}.

The order parameters are distinguished by whether they are even or odd under the in-plane 2-fold rotations ($2_{\rm 100}$, $2_{\rm 010}$ or $2_{\rm 110}$) and even or odd under inversion ($\bar{1}$). In addition, order parameters at F or L double the unit cell along specific directions depending on which point in the star the order parameter lies at. For example, the order parameter located at $\mathbf{Q}_{\rm F_1}=(0,\tfrac{1}{2},\tfrac{1}{2})$ would double the rhombohedral unit cell along the $b$- and $c$-directions. As a consequence of the little group and the structure of the stars, there are four space group irreps at both the F and L points. These are labeled by $F_{i}^{\pm}$ and $L_{i}^{\pm}$, where $i=1,2$ labels the number of the irrep and $\pm$ denotes whether it is even or odd under inversion, where the inversion center is placed at the origin of the conventional cell [Fig.~\ref{fig:r3m_crystal_and_bz}(a)] in between two kagome layers. Throughout, we use normal font to refer to points in the Brillouin zone and italic font to refer to irreps. As there are three points in the star of both F and L, all the space group irreps are three-dimensional. In contrast, the little group at T is D$_{3d}$ and the space group irreps are labeled by $T_i^{\pm}$ with $i=1,\ldots 4$, where the first two are one-dimensional and the last two are two-dimensional. The Landau free energy can be constructed by taking products between the space group irreps and keeping only the terms transforming as the trivial representation $\Gamma_1^+$, i.e. a scalar. This approach is implemented in the INVARIANTS toolkit of the ISOTROPY software suite~\cite{Hatch2003INVARIANTS,StokesInvariants}.

We consider two general cases for the coupling between order parameters at F and L. For the first case (i), the order parameter at F transforms as $F_1^+$. This has two consequences. It allows the three order parameters at F, deriving from the three points in the star of F, to couple \emph{via} a trilinear term. Moreover, it allows the F and L order parameters to couple through a separate trilinear term, regardless of which irrep the L order parameter transforms as.
Consequently, coexistence of F and L orders in this case is favored and their critical temperatures can even be increased. In case (ii), the order parameter at F transforms as $F_1^-$, $F_2^+$, or $F_2^-$, in which case no trilinear terms are allowed, and the F and L order parameters only couple at quartic order. This implies that coexistence is less robust and generally depends on specific material details.

The analysis of case (i) is more complicated due to the aforementioned trilinear terms. In Table~\ref{tab:subgroups} we depict the atomic displacements of an order parameter transforming as $F_1^+$ and, for concreteness, an order parameter transforming as $L_2^-$. This resembles the situation relevant for hole-doped \RhTlS{}, discussed in Sec.~\ref{sec:unstable_compounds}.
Since the space group irreps are three-dimensional, each order parameter will have three components with displacement patterns related to those shown in Table~\ref{tab:subgroups} by symmetry. Denoting the $F_1^+$ order parameter by $\mathbf{F}=(F_1, F_2, F_3)$ we find
\begin{align}
    \mathcal{F}_{F_1^+} &= \frac{\alpha_F}{2}(T-T_{F}) \mathbf{F}^2 + \frac{\gamma_F}{3} F_1 F_2 F_3 + \frac{u_F}{4} \mathbf{F}^4 \nonumber
    \\
    &+ \frac{\lambda_F}{4} (F_1^2 F_2^2 + F_1^2 F_3^2 + F_2^2 F_3^2)\,,
\end{align}
where $T$ denotes the temperature and $T_F$ signifies the bare temperature of the second-order phase transition to the ordered phase, i.e., without including any higher-order terms. At this point, the remaining parameters are phenomenological but they can be obtained from, e.g., a microscopic calculation. We emphasize that $F_i$ in $\mathbf{F}=(F_1, F_2, F_3)$ do not refer to different irreps, but instead, they refer to the three components of the same irrep $F_1^+$. The phase diagram of this free energy has been widely studied in the context of the \AVS{} kagome metals~\cite{Park2021Electronic,Lin2021Complex,Christensen2021Theory}. Importantly, a trilinear term is allowed which promotes simultaneous ordering of all three order parameter components and induces a first-order transition already at the mean-field level. As a consequence, the preferred ground states are those where all three components of $\mathbf{F}$ become non-zero. Borrowing terminology from the transition-metal dichalcogenides and the \AVS{} kagome metals~\cite{Wilson1969The,Jiang2021Unconventional,Park2021Electronic,Tan2021Charge}, we refer to these as either trihexagonal or Star-of-David phases, depending on the relative signs of the components of $\mathbf{F}$. The transition to either of these phases is first-order and the critical temperature is
\begin{equation}
    T_{3F} = \frac{2\gamma_F^2}{81\alpha_F (3u_F+\lambda_F)} + T_F\,,
\end{equation}
where $T_{3F}$ denotes the transition temperature of a phase where all three components of the irrep are condensed. Evidently, for $\gamma_{F}\neq 0$, $T_{3F}>T_{F}$, implying that the trihexagonal or Star-of-David configurations are favored over the striped configurations for which only a single order parameter condenses~\cite{Christensen2021Theory}. Note that we take $u_{F}>0$ and $3u_{F}+\lambda_{F}>0$ to ensure a bounded free energy. While the striped configurations break the rotational symmetry as described in Table~\ref{tab:subgroups} and doubles the unit cell in two directions, the trihexagonal and Star-of-David configurations do not break any point group symmetries, but only double the unit cell in all $\langle 100\rangle$ directions. This is similar to the case in the $P6/mmm$ kagome materials. 

In similar fashion, denoting an order parameter transforming as $L_2^-$ by $\mathbf{L}=(L_1,L_2,L_3)$, we find the free energy
\begin{align}
    \mathcal{F}_{L} &= \frac{\alpha_L}{2}(T-T_{L}) \mathbf{L}^2 + \frac{u_L}{4} \mathbf{L}^4 \nonumber
    \\
    &+ \frac{\lambda_L}{4} (L_1^2 L_2^2 + L_1^2 L_3^2 + L_2^2 L_3^2)\,.
\end{align}
Note that this free energy is the same independent of which irrep $\mathbf{L}$ transforms as. No trilinear term is allowed in this case and, depending on the sign of $\lambda_{L}$, condensation of either a single ($\lambda_{L}>0$) or all three components ($\lambda_{L}<0$) of $\mathbf{L}$ is favored, resulting in two possible phases. Their free energies are
\begin{align}
        \mathcal{F}_{1L} &= -\frac{\alpha_L^2(T-T_L)^2}{4u_L}\,, \\
        \mathcal{F}_{3L} &= -\frac{\alpha_L^2(T-T_L)^2}{4u_L + \frac{4}{3}\lambda_L}\,,
\end{align}
respectively, demonstrating that for $\lambda_{L}>0$, the $3L$ phase is favored, while for $\lambda_{L}<0$ the $1L$ phase is chosen. Note that, as above, we assume $u_{L}>0$ and $3u_{L}+\lambda_{L}>0$ to ensure a bounded free energy. The $1L$ order doubles the unit cell along one $\langle 100\rangle$ direction, while the $3L$ phase doubles the unit cell in all $\langle 100\rangle$ directions.

The free energy also allows for trilinear couplings between $\mathbf{F}$ and $\mathbf{L}$, which favors coexistence of the two orders~\cite{Park2021Electronic,Christensen2021Theory}:
\begin{align}
    & \mathcal{F}_{F_{1}^{+}L} = \frac{\gamma_{FL}}{3} (F_1 L_2 L_3 + L_1 F_2 L_3 + L_1 L_2 F_3) \nonumber
    \\
    &+ \frac{\lambda_{FL}^{(1)}}{4} (F_1 F_2 L_1 L_2 + F_1 F_3 L_1 L_3 + F_2 F_3 L_2 L_3) \nonumber
    \\
    &+ \frac{\lambda_{FL}^{(2)}}{4} (F_1^2 L_1^2 + F_2^2 L_2^2 + F_3^2 L_3^2) + \frac{\lambda_{FL}^{(3)}}{4} \mathbf{F}^2 \mathbf{L}^2\,.
\end{align}
A detailed study of this case is presented in, e.g., Ref.~\cite{Christensen2021Theory}. The main conclusions are that two phases dominate the phase diagram. These are a 3F+3L phase, in which all components of $\mathbf{F}$ and $\mathbf{L}$ condense, and a 1F+2L phase in which a single component of $\mathbf{F}$ and two components of $\mathbf{L}$ condense. In the former case, the space group remains $R\bar{3}m$ although the unit cell is doubled along all directions, while in the latter case, the order breaks rotational symmetry resulting in the space group $C2/m$ while also doubling the unit cell in all directions. These results are summarized in Table~\ref{tab:subgroups}. The phenomenology of the $R\bar{3}m$ space group exhibits a large variety of possible structural transitions which can occur as a result of a single unstable mode.

In case (ii), where the order parameter at F transforms as either $F_1^-$, $F_2^+$, or $F_2^-$, the free energy is simpler as trilinear terms are not allowed. We find
\begin{align}
    \mathcal{F}_{\slashed{F_1^+}} &= \frac{\alpha_F}{2}(T-T_{F}) \mathbf{F}^2 + \frac{u_F}{4} \mathbf{F}^4 \nonumber
    \\
    &+ \frac{\lambda_F}{4} (F_1^2 F_2^2 + F_1^2 F_3^2 + F_2^2 F_3^2)\,,
\end{align}
and
\begin{align}
    & \mathcal{F}_{\slashed{F_{1}^{+}}L} = \frac{\lambda_{FL}^{(1)}}{4} (F_1 F_2 L_1 L_2 + F_1 F_3 L_1 L_3 + F_2 F_3 L_2 L_3) \nonumber
    \\
    &+ \frac{\lambda_{FL}^{(2)}}{4} (F_1^2 L_1^2 + F_2^2 L_2^2 + F_3^2 L_3^2) + \frac{\lambda_{FL}^{(3)}}{4} F^2 L^2\,.
\end{align}
In this case, single components of the F order parameter can condense even in the absence of fine-tuning. However, since the trilinear terms are absent, whether or not states at F and L will coexist depend on specific details and must be decided on a case-by-case basis. This is in contrast to case (i), where the trilinear terms favor coexistence between the orders. Case (ii) is relevant, e.g., for hole-doped or pressurized \PdSnSe{} discussed in Sec.~\ref{sec:unstable_compounds}. This compound has a low-lying $F_{2}^+$ mode while an $L_1^+$ mode becomes unstable.

\section{First-principles calculations}\label{sec:first_principles}

To assess the role of the electronic degrees of freedom in potential lattice instabilities, we employ first-principles calculations to predict the electronic and crystal structures. Ground state electronic structures are obtained using density-functional theory (DFT) and phonon spectra are evaluated using density-functional perturbation theory (DFPT). Calculations are carried out using Abinit 9.10.1~\cite{Gonze2020, Romero2020Abinit, Gonze1997a}, within the Perdew-Burke-Ernzerhof approximation for the solids (PBEsol)~\cite{Perdew2008Restoring} and using a basis of projector augmented waves (PAW) with Jollet-Torrent-Holzwarth PAW type pseudo-potentials obtained from PseudoDojo~\cite{Jollet2014Generation, Torrent2008Implementation, Vansetten2018}. 

While the compounds studied contain Rh and Pd ions, we do not include on-site Hubbard corrections (DFT+$U$ or DFT combined with dynamical mean-field theory). Both elements are $4d$ transition metals, for which strong local correlations are generally not expected. Moreover, the calculated DFT band structures exhibit a substantial density of states at the Fermi level, implying efficient screening of on-site Coulomb interactions by itinerant bands~\cite{Paul2019}. Consequently, any physically reasonable choice of $U$ is expected to produce only minor quantitative modifications and should not affect the qualitative conclusions of this work.

We focus on the family of $M_3A_2$Ch$_2$ shandite materials, where $M$ denotes a transition metal, $A$ is a post-transition metal and $\mathrm{Ch}$ is a chalcogenide. The transition metal, $M$, can be Rh, Pd, Ni, or Co. Variants with Co or Ni have been the topic of many previous studies~\cite{Gutlich1999The,Aziz2016Electron,Liu2018Giant,Morali2019Fermi-arc,Yanagi2021First-principles,Zhang2022Endless} and we focus on the Rh and Pd realizations, which have received less attention (see, e.g., Refs.~\cite{Seidlmayer2010Half,Yu2020Pressure-induced,Basak2025Dynamical} for recent work). Hence, we calculate the ground state properties for all combinations of $M_3A_2$Ch$_2$ with $M =\mathrm{Rh, Pd}$ and $A= \mathrm{In, Pb, Sn, Tl, Bi}$ while for the chalcogen we consider either S or Se. In the shandites, the kagome layers are stacked at a $\pi/3$ angle through one of the translations in Eq.~\eqref{eq:transl_rh}, resulting in the unit cell shown in Fig.~\ref{fig:r3m_crystal_and_bz}(a) and the space group $R\bar{3}m$, as discussed in Sec.~\ref{sec:crystal_structure}. For each material, we calculate the electronic properties and the lattice response using the fully relaxed primitive rhombohedral unit cell, where the structure is deemed relaxed when the modulus of the forces on the atoms are smaller than 1 meV/\AA. We use a cutoff energy for the plane-wave basis of 650 eV, a $16 \times 16 \times 16$ $k$-point grid for the shandite structure and a $8 \times 8 \times 8$ $k$-point grid for the parkerite structures. Moreover, we converge the ground state Kohn-Sham wavefunctions until the residue squared is less than $10^{-20}$ eV$^2$, such that these can be used to calculate the dynamical matrix at first-order in perturbation theory with DFPT, for which we use a $4 \times 4 \times 4$ $q$-point grid and a tolerance of $10^{-12}$ eV in the energy difference as a convergence criterion. For most calculations we use a Gaussian smearing with an electronic smearing temperature of 1 meV, unless stated otherwise.

We report the relaxed lattice parameters in Tab.~\ref{tab:lattice_parameters} and include a comparison to experimentally measured lattice parameters in the cases where the compound has been synthesized. Additionally, Table~\ref{tab:lattice_parameters} also include comparisons of the total energies per formula unit of the $R\bar{3}m$ shandite phase and the parkerite phases with either monoclinic $C2/m$ or cubic $I2_{1}3$ structure. There is reasonable agreement between DFT-predicted and experimental parameters, and especially in the thermodynamic stability of the shandite phase. More importantly, the energy difference between different phases is at most of the order of a few tens of meV per atom in all cases. This indicates that, in principle, either metastable phase may be accessible using the right high-temperature synthesis protocol. For this reason, we include \RhBiS{}, \RhBiSe{}, \PdPbS{}, \PdBiS{}, and \PdBiSe{} in our study as well, even though they are reported only in the parkerite structure so far.

While many of the materials we consider host $4d$ transition metals which can display some spin-orbit coupling (SOC) effects in compounds with narrow bands, the DFT band structures discussed below indicate that the bandwidth is large enough in shandites that no significant SOC-induced changes are expected. In order to confirm this expectation, we present DFT results with SOC effects included in Appendix~\ref{app:SOC}. These results demonstrate that the effect of SOC grows with increasing atomic number of the post-transition metal ion, indicating that the effect primarily arises from the $p$-orbitals of these atoms. Additionally, the impact of SOC on the energy differences between the rhombohedral $R\bar{3}m$, the monoclinic $C2/m$, and the cubic $I2_{1}3$ lattices is illustrated by the numbers in the parentheses in the last two columns of Table~\ref{tab:lattice_parameters}. Despite the impact of SOC being small overall, on the order of $\sim 10$ meV per formula unit, this has an effect on \PdSnSe{} since the energy differences between the different crystal structures are small in this case.

\begin{table*}[]
    \centering
    \begin{tabular}{C{0.125\linewidth}|C{0.075\linewidth}C{0.075\linewidth}C{0.125\linewidth}C{0.125\linewidth}C{0.21\linewidth}C{0.21\linewidth}}
    \toprule
    Composition & $a$ [\AA] & $\alpha$ [$^{\circ}$] & Exp. $a$ [\AA] & Exp. $\alpha$ [$^{\circ}$] & $E_S-E_{P_m}$ [$\,$eV/f.u.] & $E_S-E_{P_c}$ [$\,$eV/f.u.] \\
    \hline \hline
    \RhInS{} & 5.549 & 60.47 & 5.571$^{\ast}$~\cite{Natarajan1988Synthesis} & 60.35$^{\ast}$~\cite{Natarajan1988Synthesis} & $-$0.461 ($-$0.441) & $-$0.536 ($-$0.516)\\
    \RhInSe{} & 5.696 & 59.18 & & & $-$0.326 ($-$0.324) & $-$0.424 ($-$0.421)\\
    \RhSnS{} & 5.491 & 61.68 & 5.491~\cite{Anusca2008Neue} & 61.47~\cite{Anusca2008Neue} & $-$0.561 ($-$0.539) & $-$0.463 ($-$0.433)\\
    \RhSnSe{} & 5.663 & 59.76 & & & $-$0.264 ($-$0.251) & $-$0.293 ($-$0.278)\\
    \RhTlS{} & 5.771 & 59.67 & 5.679~\cite{Zabel1979Ternary} & 59.55~\cite{Zabel1979Ternary} & $-$0.388 ($-$0.390) & $-$0.356 ($-$0.345)\\
    \RhTlSe{} & 5.894 & 58.39 & & & $-$0.431 ($-$0.420) & $-$0.424 ($-$0.420)\\
    \RhPbS{} & 5.677 & 60.39 & 5.657~\cite{Zabel1979Ternary} & 60.55~\cite{Zabel1979Ternary} & $-$0.173 ($-$0.190) & $-$0.325 ($-$0.338)\\
    \RhPbSe{} & 5.845 & 58.71 & 5.722$^{\ast}$~\cite{Natarajan1988Synthesis} & 60.90$^{\ast}$~\cite{Natarajan1988Synthesis} & $-$0.187 ($-$0.200) & $-$0.335 ($-$0.336)\\
    \RhBiS{} & 5.638 & 62.61  & \multicolumn{2}{c}{\centering Parkerite, $C2/m$~\cite{Range1983Neue}} & +0.370 (+0.364) & +0.248 (+0.240)\\
    \RhBiSe{} & 5.811 & 60.34 & \multicolumn{2}{c}{\centering Parkerite, $C2/m$~\cite{Sakamoto2007Charge-density-wave}} & +0.351 (+0.370) & +0.302 (+0.288)\\
    \bottomrule
    \PdInS{} & 5.620 & 61.31 & & & $-$0.377 ($-$0.359) & $-$0.201 ($-$0.184)\\
    \PdInSe{} & 5.805 & 58.72 & & & $-$0.267 ($-$0.263) & $-$0.194 ($-$0.190)\\
    \PdSnS{} & 5.597 & 63.10 & & & $-$0.117 ($-$0.102) & $-$0.086 ($-$0.070)\\
    \PdSnSe{} & 5.805 & 59.82 & & & +0.046 (+0.006) & +0.037 ($-$0.004) \\
    \PdTlS{} & 5.846 & 60.27 & 5.748~\cite{Zabel1979Ternary} & 60.52~\cite{Zabel1979Ternary} & $-$0.232 ($-$0.256) & $-$0.247 ($-$0.235)\\
    \PdTlSe{} & 6.024 & 58.03 & 5.949~\cite{Zabel1979Ternary} & 58.06~\cite{Zabel1979Ternary} & $-$0.312 ($-$0.338) & $-$0.337 ($-$0.323)\\
    \PdPbS{} & 5.758 & 62.48 & \multicolumn{2}{c}{\centering Parkerite, $C2/m$~\cite{Barkov2002Laflammeite}} & $-$0.060 ($-$0.057) & $-$0.041 ($-$0.039)\\
    \PdPbSe{} & 5.940 & 59.71 & 5.944~\cite{Zabel1979Ternary} & 59.44~\cite{Zabel1979Ternary} & $-$0.043 ($-$0.039) & $-$0.064 ($-$0.061)\\
    \PdBiS{} & 5.699 & 64.47 & \multicolumn{2}{c}{\centering Cubic parkerite, $I2_1 3$~\cite{Weihrich2007Structure}} & +0.248 (+0.272) & +0.321 (+0.306)\\
    \PdBiSe{} & 5.891 & 61.14 & \multicolumn{2}{c}{\centering Parkerite, $C2/m$~\cite{Range1983Neue}} & +0.298 (+0.297) & +0.302 (+0.283)\\
    \bottomrule
    \end{tabular}
    \caption{\label{tab:lattice_parameters} Lattice parameters obtained from DFT for the rhombohedral ($a=b=c$ and $\alpha=\beta=\gamma$) unit cell of the shandites $M_3A_2$Ch$_2$. The parameters were obtained by fully relaxing the unit cell and the atomic positions until the forces on the atoms were smaller than 1 meV/\AA{} in magnitude. Where possible, experimentally measured parameters are included for comparison. Asterisks (${}^{\ast}$) denotes cases where hexagonal parameters were converted to rhombohedral. The last two columns report the calculated energy differences between the shandite and parkerite structures; $E_S$ refers to the energy of the $R\bar{3}m$ shandite structure, $E_{P_m}$ to the energy of the monoclinic $C2/m$ parkerite structure and $E_{P_c}$ to the energy of the the cubic $I2_{1}3$ parkerite structure. The values in parentheses represent the energies obtained with SOC included.}
\end{table*}

\begin{figure}[!t]
    \centering
    \includegraphics[width=\linewidth]{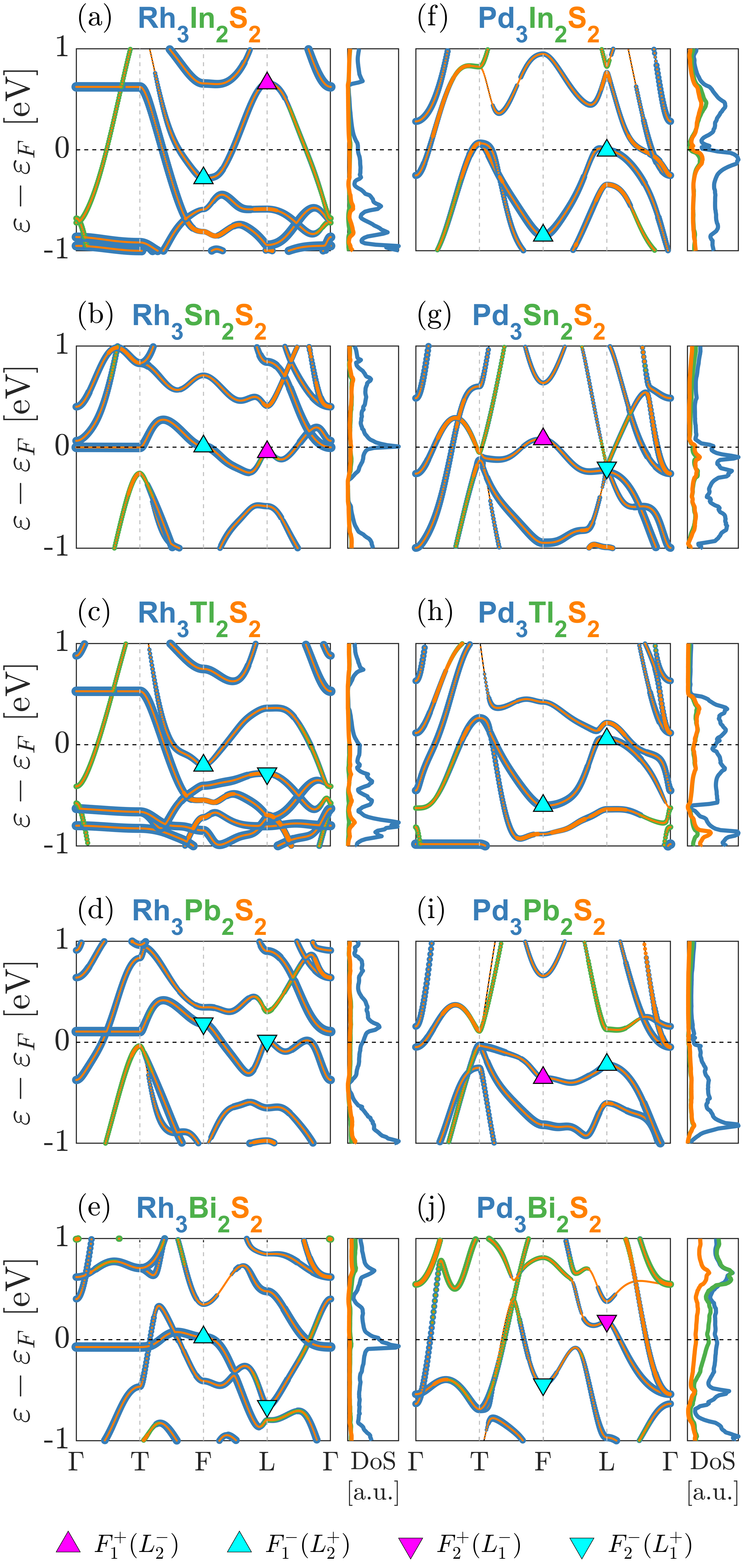}
    \caption{\label{fig:shandite_electronic_structure}Orbitally resolved electronic structures of shandite materials $M_3A_2$S$_2$ for (a)--(e) $M=$Rh and (f)--(j) $M=$Pd. Different colors correspond to different orbital weights. Blue denotes the $d$-orbitals of the $M$ atoms, green the $p$-orbitals of the $A$ atoms while orange shows the $p$-orbitals of the S atoms. The irreducible representations $F_{1,2}^\pm$ and $L_{1,2}^\pm$ of the states closer to the Fermi level are shown with magenta (+) or cyan (-) triangles pointing upwards (1) or downwards (2). The orbitally resolved electronic density of states is attached to the right of each figure.}
\end{figure}

\begin{figure}[!t]
    \centering
    \includegraphics[width=\linewidth]{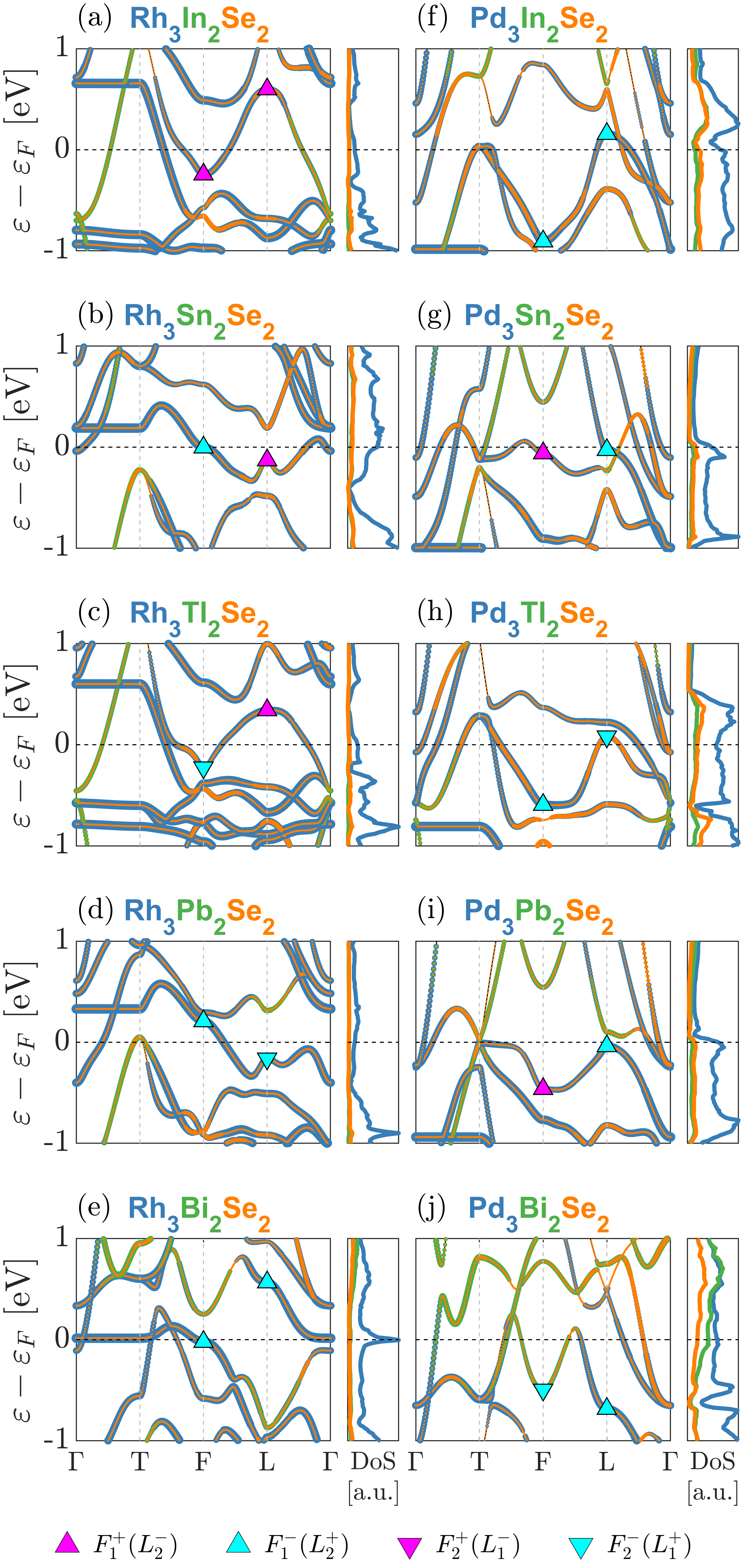}
    \caption{\label{fig:shandite_Se_electronic_structure}Orbitally resolved electronic structures of shandite materials $M_3A_2$Se$_2$ for (a)--(e) $M=$Rh and (f)--(j) $M=$Pd. Different colors correspond to different orbital weights. Blue denotes the $d$-orbitals of the $M$ atoms, green the $p$-orbitals of the $A$ atoms while orange shows the $p$-orbitals of the Se atoms. The irreducible representations $F_{1,2}^\pm$ and $L_{1,2}^\pm$ of the states closer to the Fermi level are shown with magenta (+) or cyan (-) triangles pointing upwards (1) or downwards (2). The orbitally resolved electronic density of states is attached to the right of each figure.}
\end{figure}

The electronic structures along the $\Gamma$-T-F-L-$\Gamma$ high-symmetry lines with the atomic projections are shown in Fig.~\ref{fig:shandite_electronic_structure} for the $\mathrm{Ch=S}$ case and in Fig.~\ref{fig:shandite_Se_electronic_structure} for the $\mathrm{Ch=Se}$ case. These share a number of features: The states near the Fermi level are dominated by the $d$-orbitals of Rh or Pd, with small contributions from the $p$-orbitals of the $A$ and Ch ions. We focus on the behavior of the $d$-orbitals of the $M$ ions forming the kagome layers. 

\begin{figure}[!t]
    \centering
    \includegraphics[width=.85\linewidth]{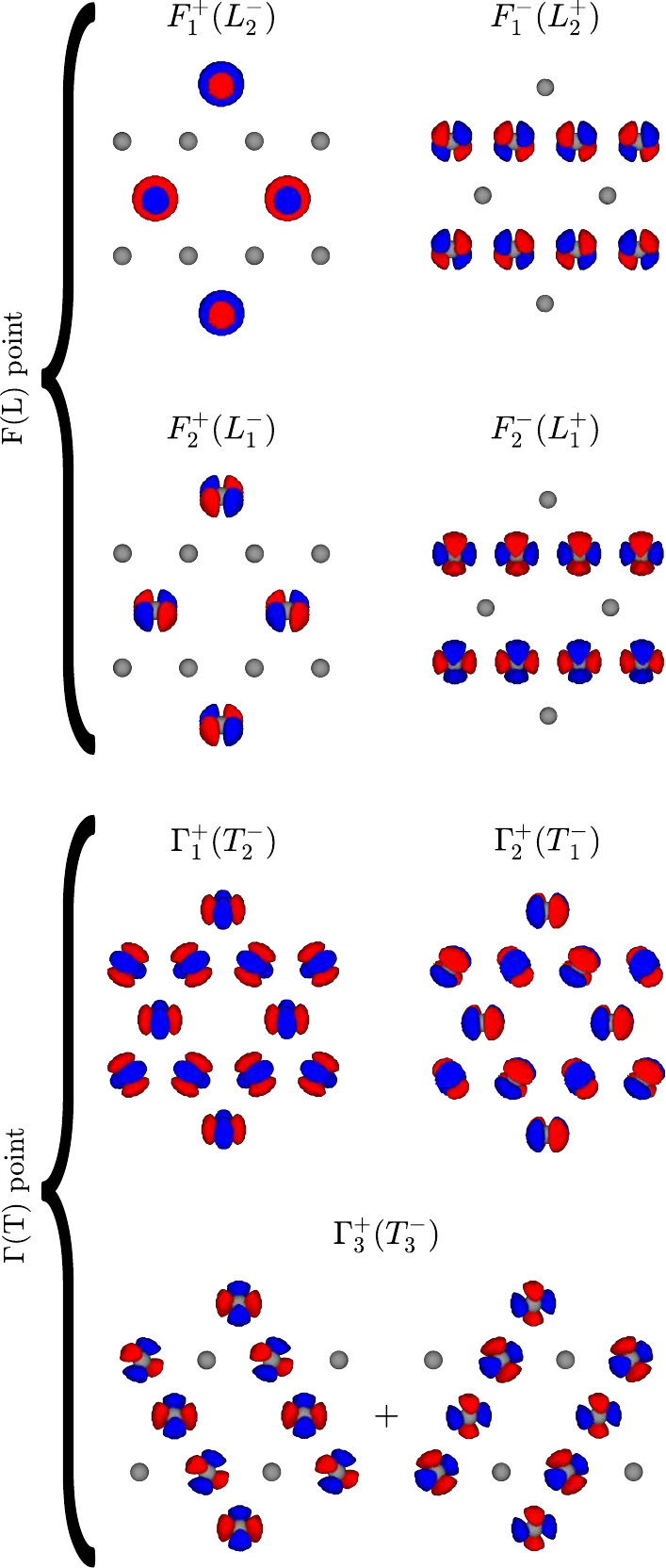}
    \caption{\label{fig:wfs_irreps} 
    Some examples of the wavefunctions induced by the $d$-orbitals of the atoms forming the kagome layers in shandites at the high-symmetry points $\Gamma$, T, F and L, and their irreps. While these wavefunctions, shown on a single layer of the conventional cell at $c=1/2$, are obtained from DFT, their forms are consistent with group theory predictions. The corresponding Wyckoff position, 9d, has local symmetry C$_{\rm 2h}$ ($2/m$) so that the d-orbitals split into the $A_g$ (even under $2_{100}$) and $B_g$ (odd under $2_{100}$) irreps. Note that the periodicity of these wavefunctions between different layers are different according to their different wavevectors. }
\end{figure}

Through the SITESYM tool~\cite{Elcoro2017Double} from the Bilbao Crystallographic Server we check the irreps that are induced by $d$-orbitals on the Wyckoff positions $9d$ for the space group $R\bar{3}m$. The site symmetry, i.e. the local symmetry of a single atom, is C$_{\rm 2h}$ ($2/m$) implying that the $d$-orbitals split into five singly degenerate states that transform as either the $A_g$ (even under the 2-fold rotation) or $B_g$ (odd under the 2-fold rotation) irreps. The particular 2-fold rotation axis is $2_{100}$, $2_{010}$ or $2_{110}$ depending on which of the three kagome atoms is considered. The induced irreps at the F, L, $\Gamma$, and T points, as well as the $\Lambda$ line ($\Lambda$ denotes the line connecting $\Gamma$ and T), are shown in Table~\ref{tab:induced_irreps}. Examples of the corresponding (real) wavefunctions are shown in Fig.~\ref{fig:wfs_irreps} for the points $\Gamma$ and F on a single layer. The T and L irreps can be readily obtained by replicating the wavefunctions with a $\pi$ phase-shift on consecutive layers.

\begin{table}[]
    \centering
    \begin{tabular}{C{0.2\columnwidth}C{0.3\columnwidth}C{0.4\linewidth}}
    \toprule
    $\mathbf{P}_{\Gamma}$ (D$_{\rm 3d}$) & Subduced irrep & M $d$-orbitals \\
    \hline \hline
    $\Gamma_1^+$ & $A_g$ & $d_{xy}+d_{x^2-y^2} + d_{z^2}$ \\
    $\Gamma_2^+$ & $B_g$ & $d_{xz} + d_{yz}$ \\
    $\Gamma_3^+$ & $A_g \oplus B_g$ & \makecell{$d_{xy}+d_{x^2-y^2} + d_{z^2}$ \\ $+d_{xz} + d_{yz}$} \\
    \midrule
    $\mathbf{P}_{\rm F}$ (C$_{\rm 2h}$) & Subduced irrep & M $d$-orbitals \\
    \hline \hline
    $F_1^+$ & $A_g$ & $d_{xy}+d_{x^2-y^2} + d_{z^2}$ \\
    $F_2^+$ & $B_g$ & $d_{xz} + d_{yz}$ \\
    $F_1^-$ & $A_g \oplus B_g$ & \makecell{$d_{xy}+d_{x^2-y^2} + d_{z^2}$ \\ $+d_{xz} + d_{yz}$} \\
    $F_2^-$ & $A_g \oplus B_g$ & \makecell{$d_{xy}+d_{x^2-y^2} + d_{z^2}$ \\ $+d_{xz} + d_{yz}$} \\
    \midrule
    $\mathbf{P}_{\rm L}$ (C$_{\rm 2h}$) & Subduced irrep & M $d$-orbitals \\
    \hline \hline
    $L_1^+$ & $A_g \oplus B_g$ & \makecell{$d_{xy}+d_{x^2-y^2} + d_{z^2}$ \\ $+d_{xz} + d_{yz}$} \\
    $L_2^+$ & $A_g \oplus B_g$ & \makecell{$d_{xy}+d_{x^2-y^2} + d_{z^2}$ \\ $+d_{xz} + d_{yz}$} \\
    $L_1^-$ & $B_g$ & $d_{xz} + d_{yz}$ \\
    $L_2^-$ & $A_g$ & $d_{xy}+d_{x^2-y^2} + d_{z^2}$ \\
    \midrule
    $\mathbf{P}_{\rm T}$ (D$_{\rm 3d}$) & Subduced irrep & M $d$-orbitals \\
    \hline \hline
    $T_1^-$ & $B_g$ & $d_{xz} + d_{yz}$ \\
    $T_2^-$ & $A_g$ & $d_{xy}+d_{x^2-y^2} + d_{z^2}$ \\
    $T_3^-$ & $A_g \oplus B_g$ & \makecell{$d_{xy}+d_{x^2-y^2} + d_{z^2}$ \\ $+d_{xz} + d_{yz}$} \\
    \midrule
    $\mathbf{P}_{\Lambda}$ (C$_{\rm 3v}$) & Subduced irrep & M $d$-orbitals \\
    \hline \hline
    $\Lambda_1$ & $A_g$ & $d_{xy}+d_{x^2-y^2} + d_{z^2}$ \\
    $\Lambda_2$ & $B_g$ & $d_{xz} + d_{yz}$ \\
    $\Lambda_3$ & $A_g \oplus B_g$ & \makecell{$d_{xy}+d_{x^2-y^2} + d_{z^2}$ \\ $+d_{xz} + d_{yz}$} \\
    \bottomrule
    \end{tabular}
    \caption{\label{tab:induced_irreps} Space group irreps at the F, L, $\Gamma$, T and $\Lambda$ points of $R\bar{3}m$ and the corresponding subduced representation at C$_{2h}$ ($2/m$), the site symmetry of the Wyckoff position 9d. For $\Gamma$ only the inversion-even irreps are shown since the space group is centrosymmetric. The $\Lambda$ points are on the $\Gamma$--T line.}
\end{table}

The bands of $M_3A_2$Ch$_2$ on the $\Gamma$--T--F--L line generally come in two varieties. The first is bands that are two-fold degenerate on the $\Gamma$--T ($\Lambda$) line and derive from the $\Gamma_3^+$ irrep at $\Gamma$, the $T_3^-$ irrep at T, and the $\Lambda_3$ irrep at a general point on the line. At the F point this splits into one of the one-dimensional $F$ irreps (here we are talking about the irreps of a \emph{single} F point of the star, hence these are one-dimensional). The second variety is bands that are one-dimensional, non-degenerate by symmetry through the whole line and thus only exhibit accidental degeneracies (if any). The least dispersive bands on the $\Lambda$ line, some of which are completely flat, derive from the $A_g$ orbitals simultaneously present on all three kagome atoms, transforming like $\Gamma_1^+$, $\Lambda_1$ and $T_2^-$. Such flat bands appear very close to the Fermi level in some of the Rh compounds, but well below it in the Pd compounds, as can be seen by contrasting the first and second column in Figs.~\ref{fig:shandite_electronic_structure} and \ref{fig:shandite_Se_electronic_structure}. The more dispersive bands on the $\Lambda$ line come from $B_g$ orbitals forming either one-dimensional $\Gamma_2^+$, $\Lambda_2$, and $T_1^-$ states or the two-dimensional $\Gamma_3^+$, $\Lambda_3$, and $T_3^-$ states. 

At the F and L points, there are only four possible states $F_{1,2}^\pm$ and $L_{1,2}^\pm$. When combining the transformation properties under the point group irreps with the translational ones imposed by the F and L points, one can infer how the Bloch states localize on specific kagome atoms within a single unit cell. As shown in Fig.~\ref{fig:wfs_irreps}, the $F_{1}^-$ ($L_{2}^+$) and $F_{2}^-$ ($L_{1}^+$) states arise from contributions across all orbitals and are distributed over two of the three atoms, a result of the `mixed' nature of the van Hove points. In contrast, $F_{1}^+$ ($L_{2}^-$) consists solely of $A_g$ orbitals localized on a single atom, while $F_{2}^+$ ($L_{1}^-$) is composed exclusively of $B_g$ orbitals, also confined to a single atom. These are examples of `pure' type van Hove points.

\begin{figure}[!]
    \centering
    \includegraphics[width=0.95\linewidth]{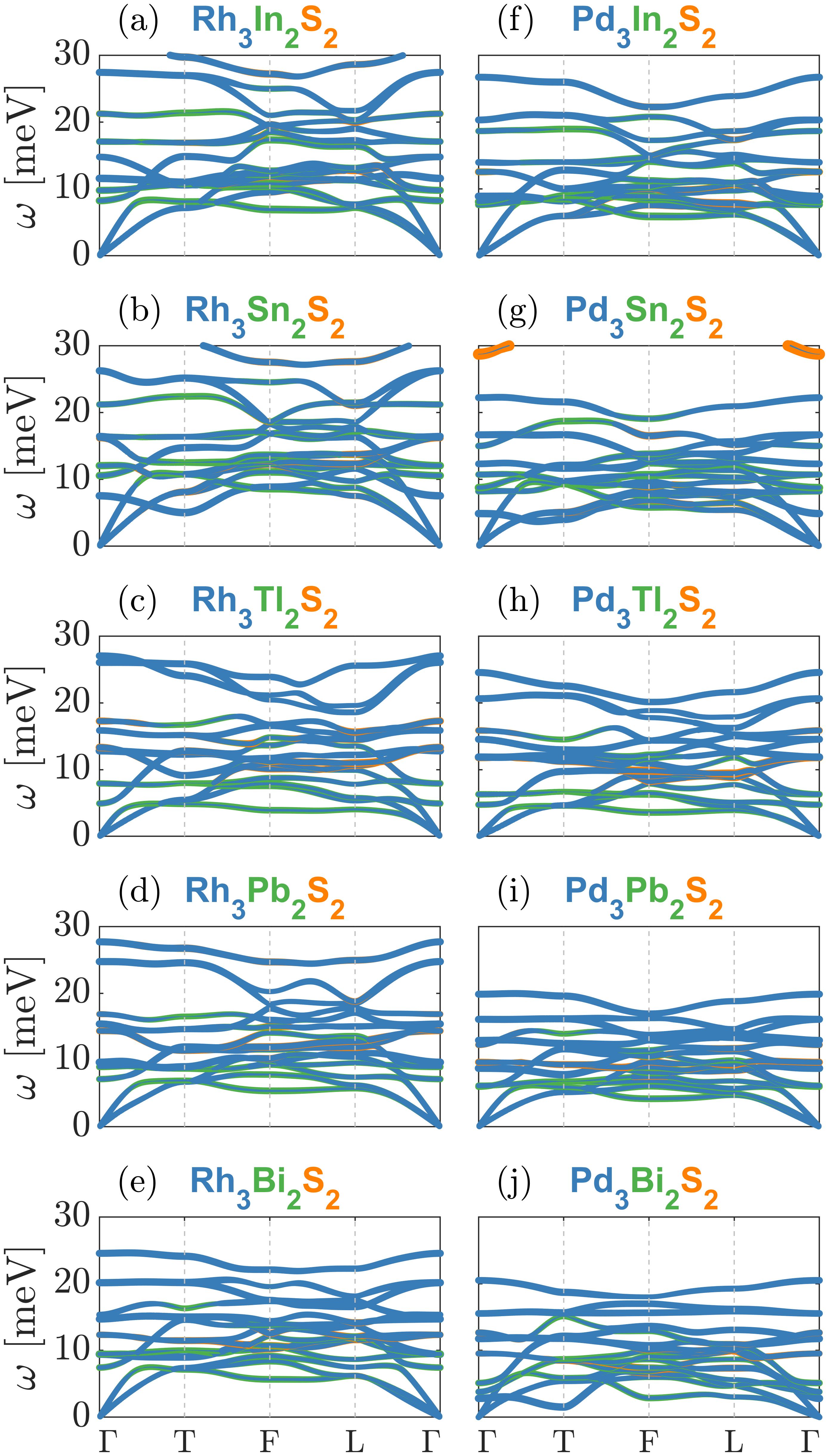}
    \caption{\label{fig:shandite_phonon_spectra} Phonon spectra of shandite materials $M_3A_2$S$_2$, with contributions from different atomic species highlighted: blue represents displacement primarily from M atoms, green corresponds to A atoms, and orange denotes S atoms. Higher energy phonons, completely dominated by the motion of the S ions, are not shown.}
\end{figure}

\begin{figure}[!t]
    \centering
    \includegraphics[width=0.95\linewidth]{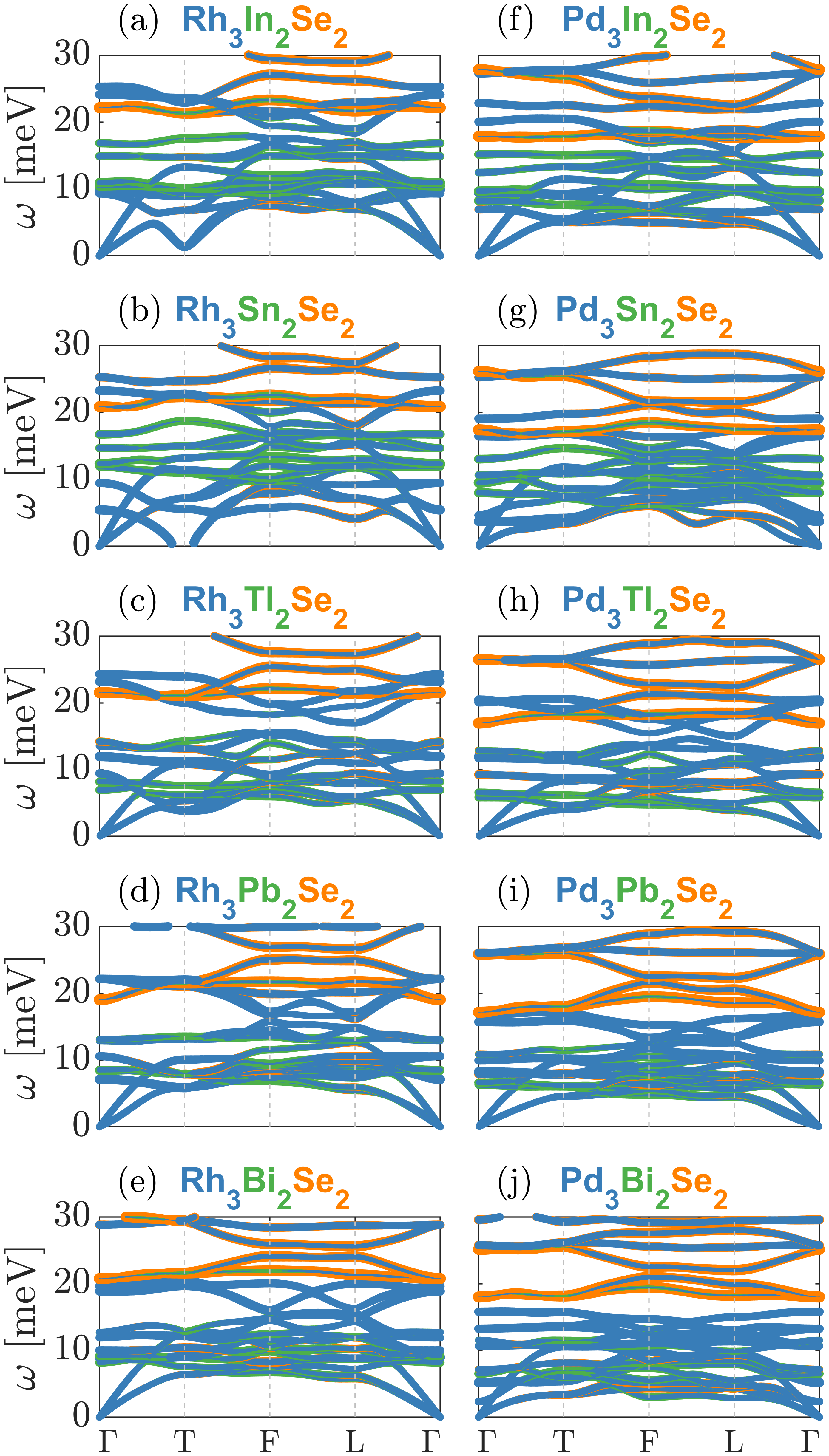}
    \caption{\label{fig:shandite_Se_phonon_spectra} Phonon spectra of shandite materials $M_3A_2$Se$_2$, with contributions from different atomic species highlighted: blue represents displacement primarily from M atoms, green corresponds to A atoms, and orange denotes Se atoms. Higher energy phonons, completely dominated by the motion of the Se ions, are not shown.}
\end{figure}

The phonon spectra of $M_3A_2$Ch$_2$ shandites exhibit distinct energy-dependent contributions from the constituent atoms. At low energies, the spectra are dominated by phonon modes primarily involving the motion of the heavier $M$ and $A$ ions, while at higher energies, the vibrations of the Ch ions become predominant, owing to their lighter masses. This behavior is illustrated in Fig.~\ref{fig:shandite_phonon_spectra} and Fig.~\ref{fig:shandite_Se_phonon_spectra}. A general trend is observed wherein certain phonon modes associated with the $M$ ions display dips at the $T$ point. Nevertheless, for most of the studied compounds, the phonon spectra remain stable indicating that no structural instabilities nor charge orders are predicted by DFT at ambient conditions. The sole exception is \RhSnSe{}, which exhibits an unstable $T_2^+$ phonon mode.
We attribute this exception to a purely structural instability, as will be explained later. Moreover, the Bi-based compounds are also found to be structurally stable, providing further indications that it may be possible to synthesize these in the shandite structure. To probe how these predictions relate to the electronic degrees of freedom, we focus on materials featuring F or L states near the Fermi level and away from the $\Gamma$--T flat bands. These criteria are met by the materials \RhTlS{}, \PdSnS{}, and \PdPbS{}, as shown in Figs.~\ref{fig:shandite_electronic_structure}(c), (g), and (i), respectively, as well as \RhSnSe{} and \PdSnSe{}, as depicted in Figs.~\ref{fig:shandite_Se_electronic_structure}(b) and (g).

Using simulated doping and simulated hydrostatic pressure, the energy of the F and L states can be tuned across the Fermi level thus allowing for the study of the potential correlation between the position of the electronic states and the appearance of any unstable phonon modes. Doping is simulated by introducing a background charge corresponding to a fraction of an electron, uniformly distributed in the unit cell, such that the charge density is $n_x(\mathbf{r})=n(\mathbf{r})+ex$, where $-e$ is the elementary charge of the electron and $x$ represents the fraction of added charge. Consequently, positive $x$ corresponds to hole doping and vice versa. We note that this approach of simulated doping is only justifiable if $x$ is kept small. After introducing the background charge, we do not perform any further structural relaxation. Pressure is simulated by imposing that the diagonal elements of the stress tensor reach a target value of $-P$, corresponding to the magnitude of the hydrostatic pressure, when relaxing the crystal structure of the material being studied. In the theoretical calculations, we apply pressures up to 50 GPa, below the 80 GPa reached in a recent experimental study~\cite{Yu2020Pressure-induced}.

\subsection{Structurally stable compounds: \PdSnS{}, \PdPbS{}, and \RhSnSe{}}\label{sec:stable_compounds}

\begin{figure*}
    \centering
    \includegraphics[width=\linewidth]{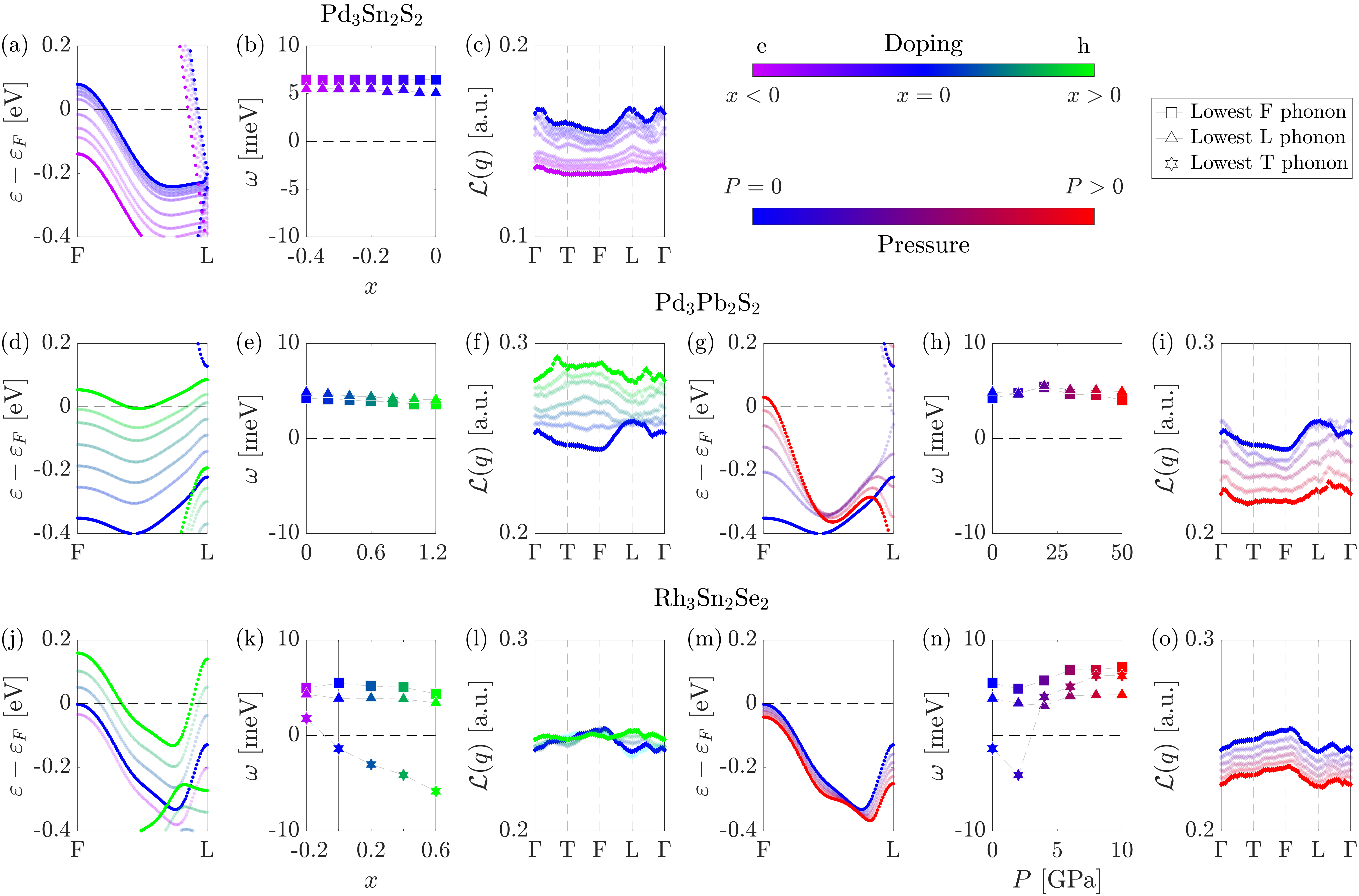}
    \caption{\label{fig:TMX_doping_pressure} Band structure, phonon frequencies and Lindhard function of (a)--(c) \PdSnS{}, (d)--(i) \PdPbS{} and (j)--(o) \RhSnSe{} under doping and pressure. Imaginary frequencies are represented as negative numbers for easier visualization. For \RhSnSe{}, the T point phonon mode is stabilized by electron doping (k), or by pressures exceeding $4$ GPa (n).}
\end{figure*}

It is instructive to first consider the cases where doping or pressure fails to induce a structural distortion. Fig.~\ref{fig:TMX_doping_pressure} shows the effects of doping and pressure on \PdSnS{}, \PdPbS{}, and \RhSnSe{}. In addition to the changes in the electronic structure, we calculate the doping-dependence of the lowest frequency phonon mode at the F and L points respectively, as well as the Lindhard function. The Lindhard function is a measure of the nesting of the electronic structure, and it is given by
\begin{equation}
    \mathcal{L}(\mathbf{q}) = \frac{1}{N^2_bN_k}\Re\left\{\sum_{nm\mathbf{k}} \frac{n_F(\varepsilon_{n\mathbf{k}+\mathbf{q}})-n_F(\varepsilon_{m\mathbf{k}})}{\varepsilon_{m\mathbf{k}}-\varepsilon_{n\mathbf{k}+\mathbf{q}}+i\eta}\right\}\,,\label{eq:lindhard}
\end{equation}
where $\varepsilon_{n\mathbf{k}}$ is the energy of band $n$ at momentum $\mathbf{k}$ and $n_F(\cdot)$ denotes the Fermi-Dirac distribution function where we use an electronic temperature of 1~meV. The expression is normalized by the number of bands $N_b^2$ and the number of $k$-points sampled $N_k$. We selected all the bands in a window of 2~eV around the Fermi level. For the artificial smearing we use $\eta \approx 10^{-3}$ meV, corresponding to the average separation of the energy eigenstates. To calculate the Lindhard function, we evaluated the eigenvalues on a finer $32\times 32 \times 32$ $k$-point grid through a one-shot non-selfconsistent DFT calculation, so that $N_k=32^3$.
In the one-band case, Eq.~\eqref{eq:lindhard} is precisely the non-interacting susceptibility. In multi-orbital systems, matrix elements quantifying the degree of overlap between the states connected by $\mathbf{q}$ also play a role. However, these are omitted here, as we focus solely on whether the electronic structure exhibits increased or decreased nesting as the electronic content is varied.

Doping moves the electronic states at F and/or L closer to the Fermi level and leads to a crossing for a specific value of $x$, particular to the material [see Figs.~\ref{fig:TMX_doping_pressure}(a), (d), and (j)]. In the case of \PdSnS{}, electron doping shifts the $F_1^+$ state closer to the Fermi level. As shown in Fig.~\ref{fig:TMX_doping_pressure}(c), the Lindhard function exhibits a relative peak at the L point that seems to be suppressed as the band is moved below the Fermi level. As seen in Fig.~\ref{fig:TMX_doping_pressure}(b), no concomitant softening of the phonon modes is observed and the material remains stable within this doping range. 

For \PdPbS{} electron doping shifts both the $F_1^+$ and $L_2^+$ states closer to the Fermi level, as shown in Fig.~\ref{fig:TMX_doping_pressure}(d). While the Lindhard function, shown in Fig.~\ref{fig:TMX_doping_pressure}(f) is enhanced slightly for large values of electron doping, with a few peaks forming, this is for values of $x$ corresponding to adding more than one electron per unit cell. This is beyond the range where the simple method we used to simulate doping  is deemed justifiable and does not result in any of the phonon modes becoming unstable, as seen in Fig.~\ref{fig:TMX_doping_pressure}(e).

\begin{figure}
    \centering
    \includegraphics[width=\linewidth]{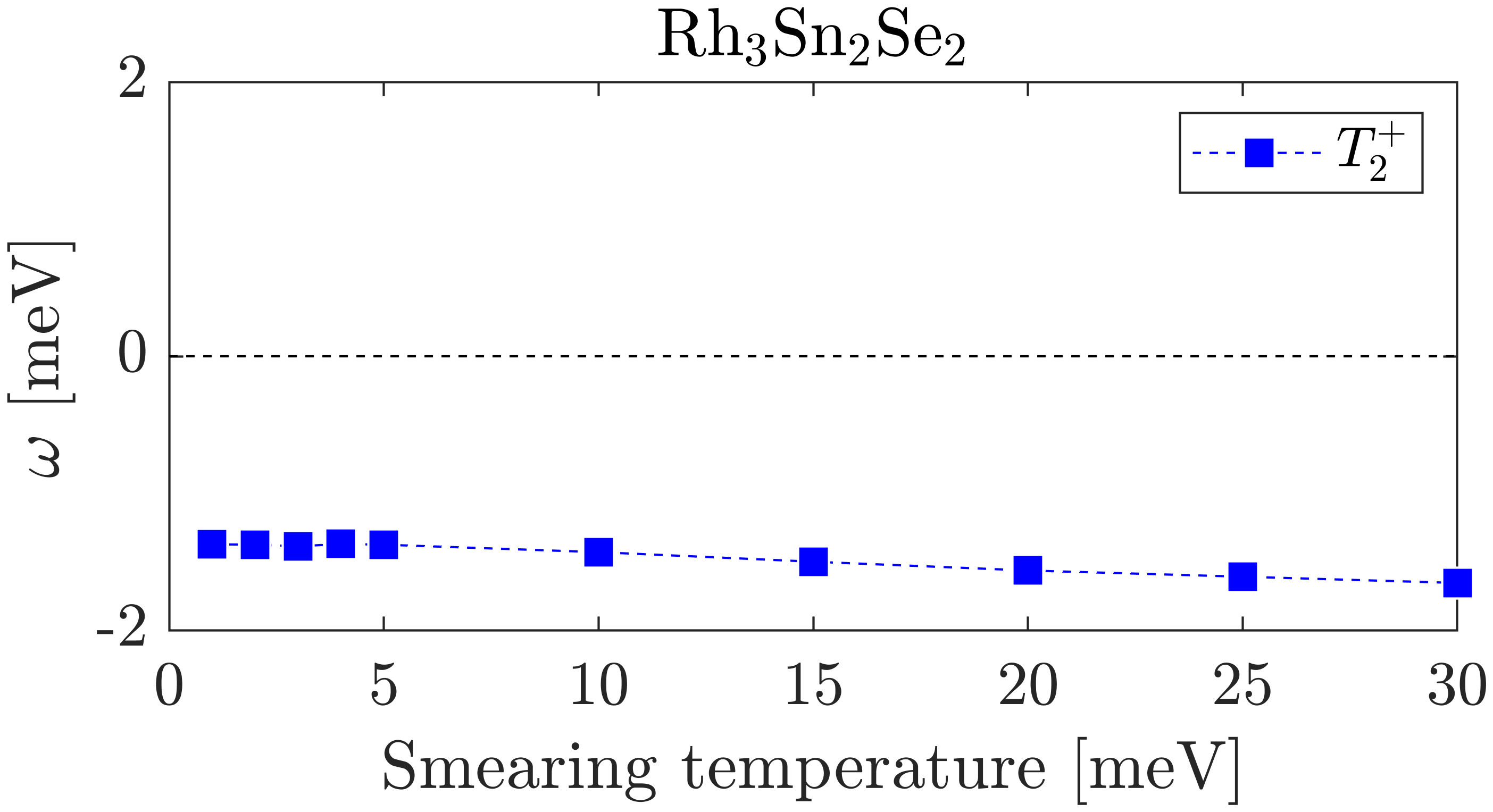}
    \caption{Dependence on the electronic smearing temperature of the unstable $T^+_2$ phonon mode of \RhSnSe{}. The nearly flat behavior indicates that the transition is driven primarily by structural degrees of freedom rather than electronic. Here, imaginary frequencies are represented by real negative numbers for easier visualization.}
    \label{fig:ElecTemp_RhSnSe}
\end{figure}

In \RhSnSe{}, the $F_1^+$ state appears at the Fermi level and any doping shifts it further away. In contrast, the $L_2^-$ state is shifted closer to the Fermi level by hole doping, as shown in  Fig.~\ref{fig:TMX_doping_pressure}(j). Thus, hole doping leads to the formation of a peak near L in the Lindhard function [Fig.~\ref{fig:TMX_doping_pressure}(l)], although the F and L point phonons remain stable throughout the doping range, as shown in Fig.~\ref{fig:TMX_doping_pressure}(k). This peak is observed for the substantially hole doped case, which is beyond the validity of the background charge method. \RhSnSe{} does exhibit an unstable $T_2^+$ phonon mode at T. However, as shown in Fig.~\ref{fig:ElecTemp_RhSnSe}, this is unaffected by the electronic smearing temperature, and we therefore attribute this to a structurally driven transition, with a negligible role played by the electrons and the electron-phonon coupling. As seen in Fig.~\ref{fig:TMX_doping_pressure}(k), a relatively small amount of electron doping also renders the $T_2^+$ phonon mode stable.

Applying hydrostatic pressure shifts the electronic states of \PdPbS{} and \RhSnSe{} closer to the Fermi level. In contrast, hydrostatic pressure on \PdSnS{} has the effect of moving the $F_1^+$ state away from the Fermi level, so we do not consider it here. In the case of \PdPbS{}, a large pressure of $P\approx42$ GPa is required to shift the $F_1^+$ state to the Fermi level [see Fig.~\ref{fig:TMX_doping_pressure}(g)]. The Lindhard function [Figs.~\ref{fig:TMX_doping_pressure}(i) and (o)] is generally suppressed and develops no new features under pressure. Moreover, the L and F point phonons remain stable under pressure [as shown in Figs.~\ref{fig:TMX_doping_pressure}(h) and (n)], and the $T_2^+$ mode of \RhSnSe{} becomes stable near $P \approx 4$ GPa, as shown in Fig.~\ref{fig:TMX_doping_pressure}(n).

We see that, despite the van Hove points in these being compounds being brought near the Fermi level by either doping or pressure, no structural instabilities emerge. This is in line with the results of Ref.~\cite{Johannes2008Fermi}, demonstrating that the electron-phonon coupling should also be sufficiently large for a coupled structural and CDW instability to appear.

\subsection{Instabilities in \RhTlS{} and \PdSnSe{}}\label{sec:unstable_compounds}

\begin{figure}
    \centering
    \includegraphics[width=0.95\linewidth]{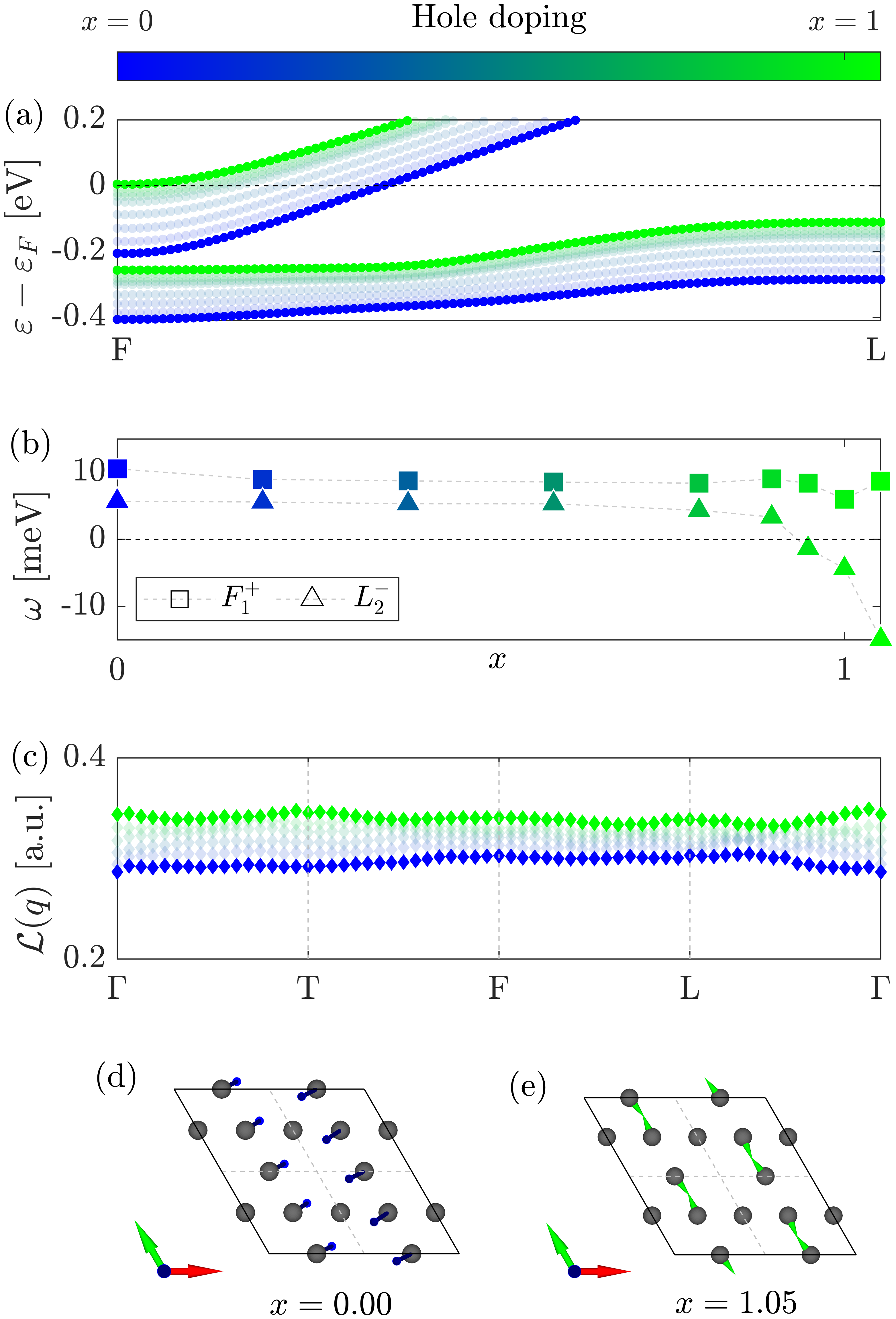}
    \caption{\label{fig:RhTlS}\RhTlS{} under simulated hole doping. (a) Electronic band structure between the F and L points as a function of doping. (b) Lowest energy phonon modes at the F and L points as a function of doping. Negative values represent imaginary frequencies. (c) Bare electronic susceptibility in the constant matrix element approximation, as a function of doping. (d) Lowest energy $L_2^-$ phonon mode at $x=0$ (stable) and at $x=1$ (unstable).}
\end{figure}

\begin{figure}
    \centering
    \includegraphics[width=\linewidth]{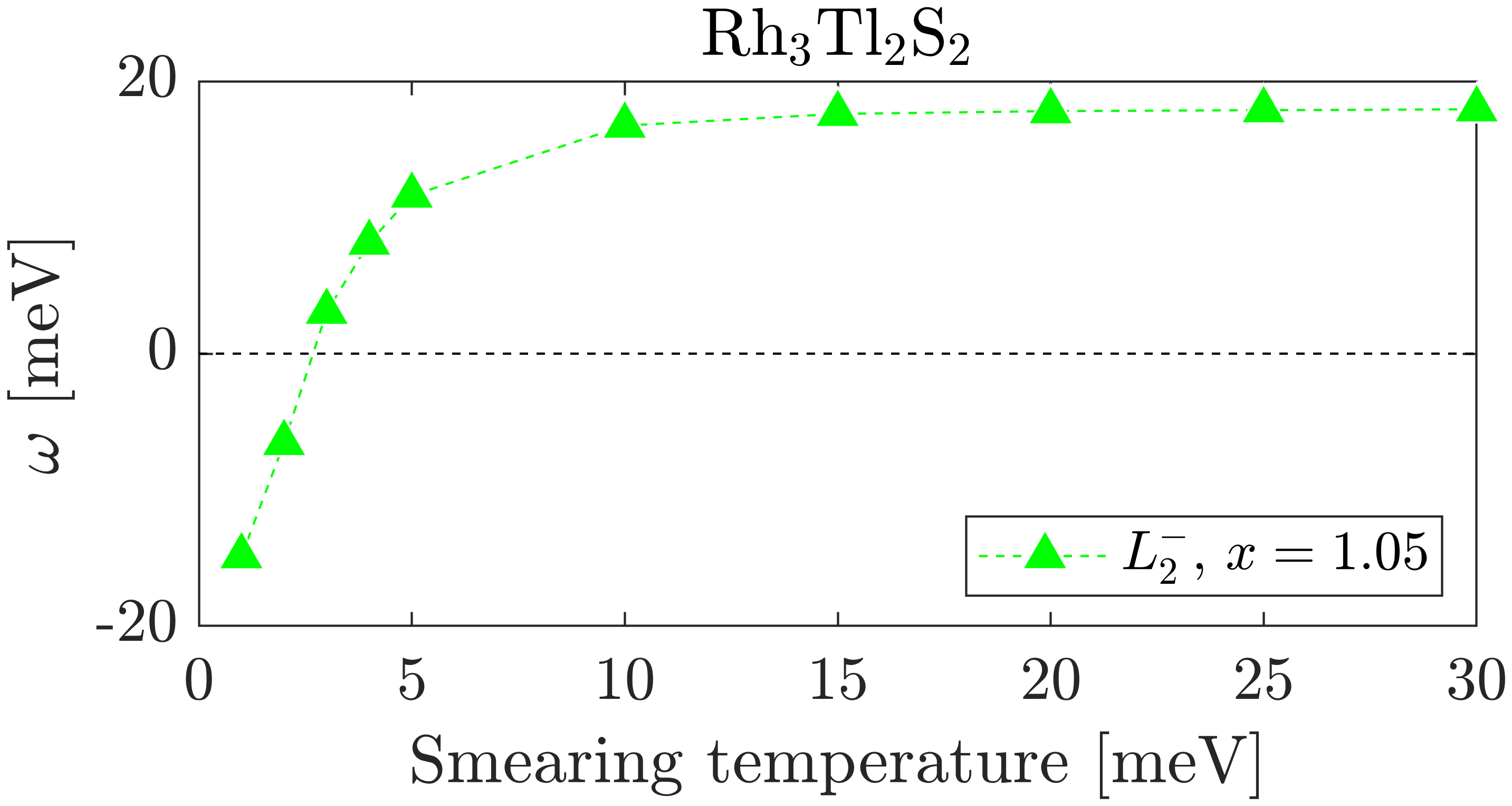}
    \caption{Dependence of the unstable phonon modes frequency on the electronic smearing temperature for the $L_2^-$ phonon mode in electron doped \RhTlS{} at $x=1.05$. The sharp dependence on electronic temperature suggests a transition driven primarily by the electronic degrees of freedom. Imaginary frequencies are represented as negative numbers for easier visualization.}
    \label{fig:ElecTemp_RhTlS}
\end{figure}

Details of the first-principles results on \RhTlS{} are presented in Fig.~\ref{fig:RhTlS}. 
The electronic structure is dominated by the Rh $d$-orbitals as shown in Fig.~\ref{fig:shandite_electronic_structure}(c), although the S $p$-orbitals are prevalent in the states near the L point. Neither the $F_1^+$ or $L_1^+$ states nearest to the Fermi level are accompanied by a peak in the density of states. Nevertheless, hole doping [Figs.~\ref{fig:RhTlS}(a) and (b)] induces a structural instability near $x = 1$, when $F_1^+$ lies at the Fermi level. Once again, this is taking the doping method beyond its realm of reliability when comparing to experiments, but the fact that a change in the Fermi level leads to soft phonon mode hints at the interplay between structural and electronic degrees of freedom in this compound, especially when combined with the fact that the instability disappears when increasing the electronic smearing temperature, as shown in Fig.~\ref{fig:ElecTemp_RhTlS}. In spite of this, the Lindhard function, shown in Fig.~\ref{fig:RhTlS}(c) is featureless and exhibits no enhancement as the structural instability is approached. We note that the smearing temperature for which the phonon becomes stable, $\sim 3$ meV in Fig.~\ref{fig:ElecTemp_RhTlS}, should not be interpreted as the transition temperature of the structural transition. As shown in Ref.~\cite{Gutierrez-Amigo2024Phonon}, anharmonic contributions will generally impact the transition temperature. Rather, the strong dependence of the phonon frequency on the electronic smearing demonstrates that the electrons, through the electron-phonon coupling, play an important role in driving the structural deformation.

The soft phonon mode is a $L_2^-$ mode associated with the distortion pattern illustrated in Fig.~\ref{fig:RhTlS}(d). When this mode is stable, it primarily involves out-of-plane motion of the Rh ions. However, as the phonon frequency becomes imaginary, the mode evolves into a pattern resembling that shown in Table~\ref{tab:subgroups}, characterized predominantly by in-plane motion of the Rh ions in a dimerized configuration. As explained in Sec.~\ref{sec:phenomenology}, the free energy is minimized by a state where either a single mode or three modes condense simultaneously. In the former case, the symmetry is lowered to $C2/m$, while in the latter the space group remains $R\bar{3}m$, although the unit cell is doubled along all three lattice directions.
Here, the lowest phonon mode at F is an $F_1^+$ mode. Hence, this corresponds to case (i) for the free energy studied in Sec.~\ref{sec:phenomenology}. Consequently, it is possible that the condensation of the $L_2^-$ order results in a secondary transition of an $F_1^+$ order. Determining whether or not this is the case would require calculating the energy of the appropriately distorted lattice as done, e.g., in Refs.~\cite{Ratcliff2021Coherent, Ritz2023Impact}  for the kagome metals.

\begin{figure}
    \centering
    \includegraphics[width=\linewidth]{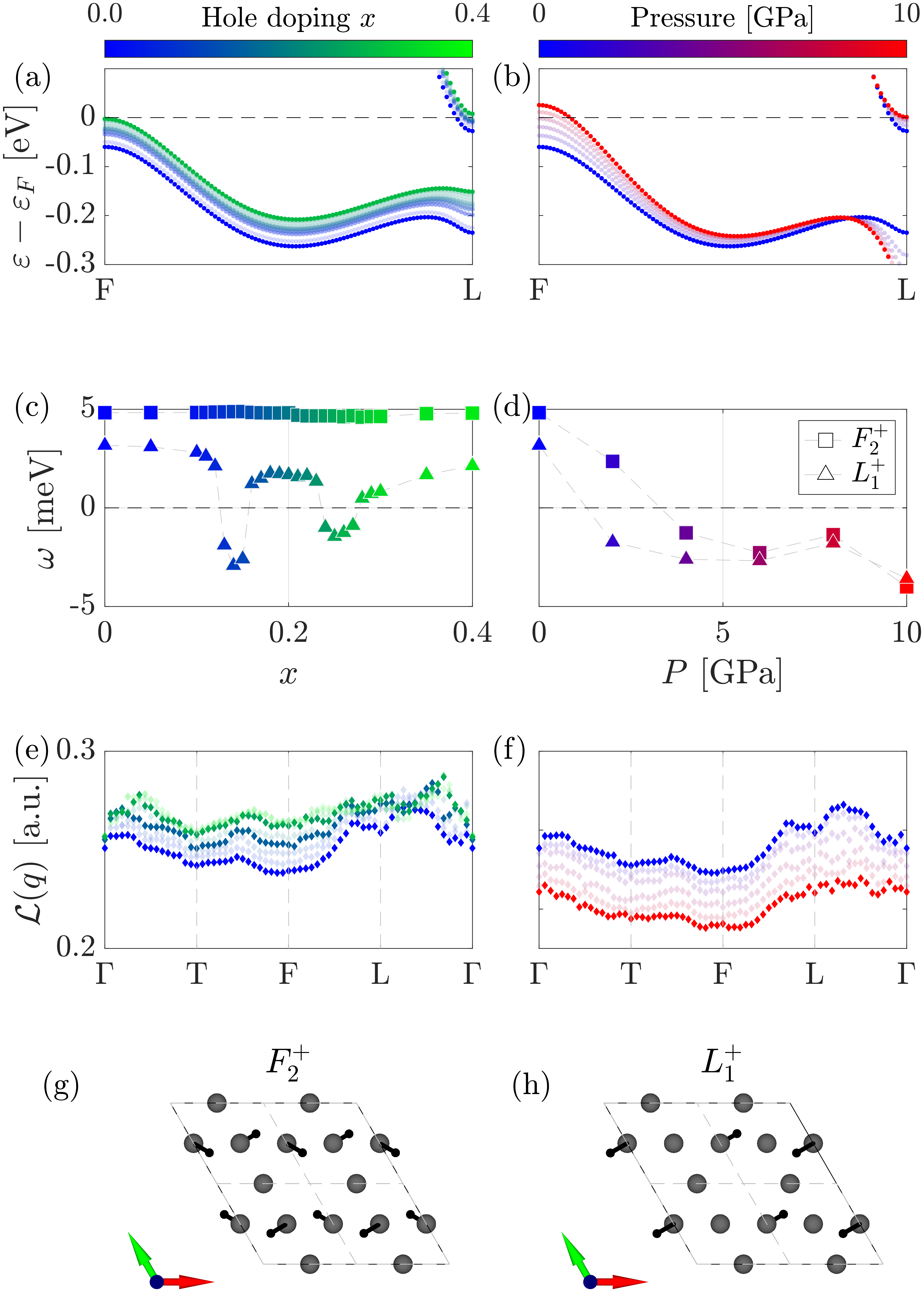}
    \caption{\label{fig:PdSnSe} \PdSnSe{} under simulated hole doping and hydrostatic pressure. (a--b) Electronic band structure in the line between the F and L point as a function of hole doping and hydrostatic pressure. (c--d) Lowest energy phonon modes at the F and L points as a function of hole doping and hydrostatic pressure. Negative values represent imaginary frequencies. (e--f) Bare electronic susceptibility in the constant matrix element approximation, as a function of hole doping and hydrostatic pressure. (g--h) Lowest energy phonon modes that become unstable under doping or pressure at the F point (g) and at the L point (h).}
\end{figure}

Like \RhTlS{}, the electronic structure of \PdSnSe{} near the Fermi level is dominated by electronic $d$-orbitals, in this case originating from Pd, as shown in Fig.~\ref{fig:shandite_Se_electronic_structure}(g). The density of states exhibits a prominent peak immediately below the Fermi level, and the electronic structure features segments of near-flat bands between F and L and between L and T. As shown in Figs.~\ref{fig:PdSnSe}(c) and (d), both hole doping and hydrostatic pressure induce structural transitions in this material. We caution that the instabilities of \PdSnSe{} are not to either of the parkerite phases, despite these being close in energy (see Table~\ref{tab:lattice_parameters}). As explained above, the parkerite phases are not reachable from the shandite phase by a deformation corresponding to a single unstable phonon mode.

In the case of hole doping, a soft acoustic mode with symmetry $L_1^+$ appears near $x \approx 0.15$ and $x \approx 0.25$, with an intermediate stable region. These instabilities occur at around the same doping level where the electronic states $F_1^+$ and $L_2^+$, respectively, cross the Fermi level. As in the previous cases, the Lindhard function remains nearly featureless and exhibits no enhancements near these doping levels. Hydrostatic pressure also leads to soft modes, with both an $L_1^+$ and an $F_2^+$ mode becoming soft near 2 GPa and 4 GPa, respectively. As with doping, these instabilities are concomitant with states at F and L crossing the Fermi level while the Lindhard function remains featureless. Differently from the case of \RhTlS{}, the two unstable phonon modes do not experience significant changes after becoming unstable, and they distort the kagome lattice as shown in Figs.~\ref{fig:PdSnSe}(g) and (h). As the low-lying mode at F transforms as $F_2^+$, this corresponds to case (ii) in Sec.~\ref{sec:phenomenology}.

\begin{figure}
    \centering
    \includegraphics[width=\linewidth]{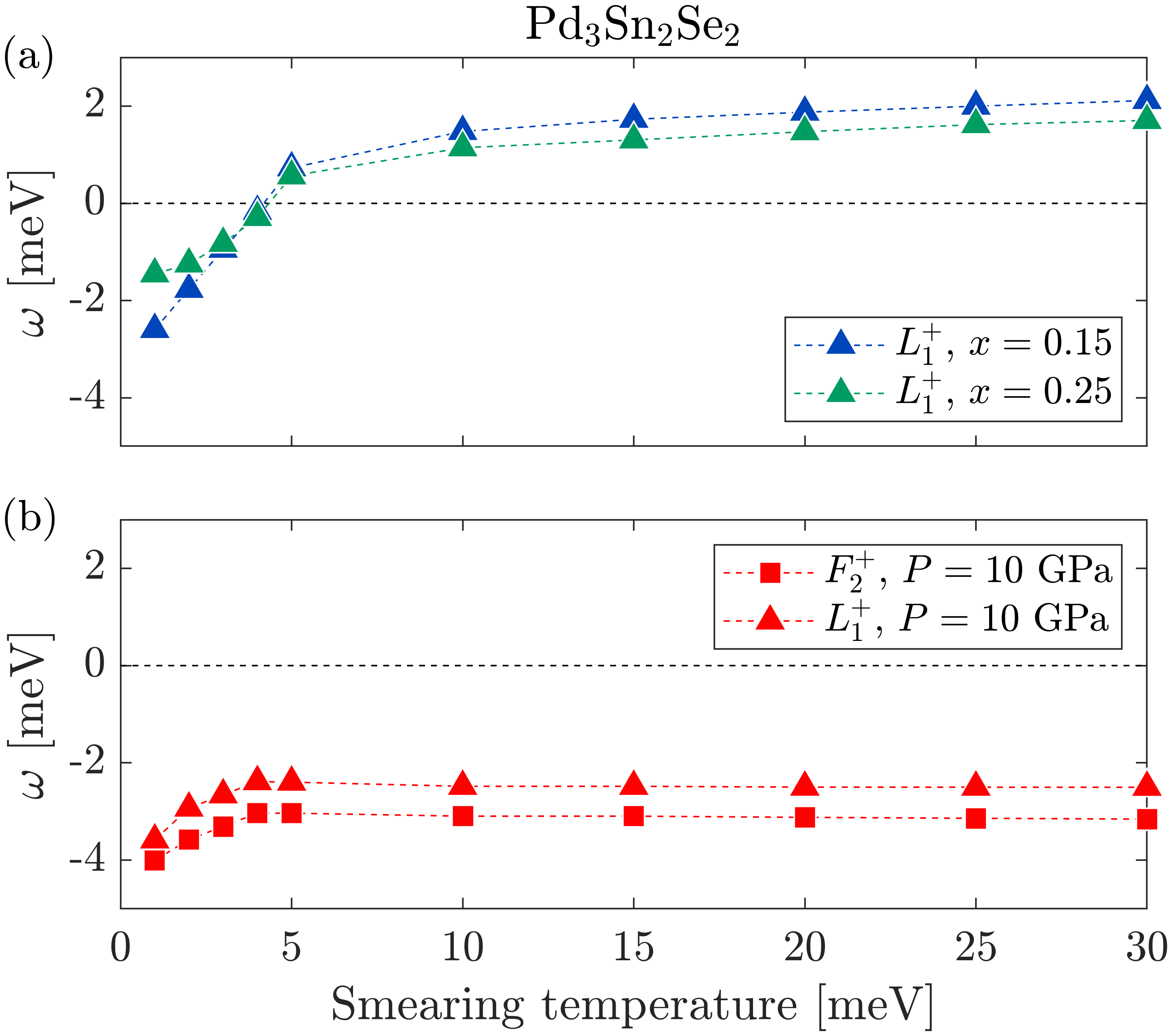}
    \caption{Dependence of the unstable phonon modes frequency on the electronic smearing temperature for (a) the $L_1^+$ phonon mode in electron doped \PdSnSe{} at $x=0.15$ and $x=0.25$, indicating a transition driven primarily by electronic degrees of freedom (b) the $L_1^+$, $F_2^+$ phonon modes in \PdSnSe{} under 10GPa of hydrostatic pressure. The flat behavior suggests that the transition is driven by the structural degrees of freedom, in contrast to the hole doped scenario. Imaginary frequencies are represented as negative numbers for easier visualization.}
    \label{fig:ElecTemp_PdSnSe}
\end{figure}

To further probe the interplay between the structural and electronic degrees of freedom, we performed the above calculations at different electronic smearing temperatures. The results are shown in Fig.~\ref{fig:ElecTemp_PdSnSe}. As the electronic smearing temperature is increased, the instability under hole doping disappears, thus indicating that the electrons themselves play a role in driving the structural instability [Fig.~\ref{fig:ElecTemp_PdSnSe}(a)], despite the fact that the Lindhard function exhibits no peaks near the relevant high symmetry points. Once again, this is in line with the results of Ref.~\cite{Johannes2008Fermi}, indicating that the electron-phonon coupling is playing an important role. The instability under pressure shows a noticeable dependence on the smearing temperature only at low temperatures [Fig.~\ref{fig:ElecTemp_PdSnSe}(b)]. This behavior contrasts with that of \RhSnSe{}, shown in Fig.~\ref{fig:ElecTemp_RhSnSe}, and suggests that hydrostatic pressure may partially stabilize the structural instability compared to the hole-doped case, possibly suggesting that a different mechanism is at play.

The $L_1^+$ phonon mode is illustrated in Fig.~\ref{fig:PdSnSe}(h) and the main distinction from an F-point phonon mode is the additional out-of-phase modulations between neighboring kagome layers. As discussed in Sec.~\ref{sec:phenomenology}, instabilities at the L point can lead to two distinct phases, depending on whether one or three L modes condense. In the case of the condensation of a single $L_1^+$ mode, the three-fold improper rotation symmetry is broken, and the space group is reduced to $C2/m$ (\#12). If three $L_1^+$ modes condense, the unit cell doubles in all lattice directions while the space group remains $R\bar{3}m$. An F-point mode can condense in addition to the L modes. This can happen in two distinct ways: either through a secondary transition as in Fig.~\ref{fig:PdSnSe}(d) resulting from a soft phonon mode or by being induced by the L order through the couplings discussed in Sec.~\ref{sec:phenomenology}. Here, the lowest phonon mode at the F point transforms as $F_2^+$, implying that there is no trilinear coupling to the condensed $L_1^+$ mode, corresponding to case (ii) discussed in Sec.~\ref{sec:phenomenology}. In this case, whether coexistence of the two orders is favored depends on the specific parameters of the free energy. To determine if an additional order condenses as a result of the $L_1^+$ mode, one should compute the energy of the different crystal structures which is beyond the scope of the current manuscript.

\section{Conclusions and Discussion}\label{sec:conclusions}

In this work, we carried out a detailed study of possible electronically driven structural phase transitions in Pd- and Rh-based kagome-layered shandites. Combining DFT and DFPT we calculated the electronic structure and the phonon spectra for 20 shandite candidate materials crystallizing in the $R\bar{3}m$ space group. The electronic structures are dominated by the Pd or Rh $d$-electrons and feature van Hove points at the F and L points in the Brillouin zone, and in many of the Rh-based compounds a flat band appears near the Fermi level between the $\Gamma$ and T points. All combinations studied save one were found to be (meta)stable, i.e., only one exhibited an imaginary frequency phonon. The unstable material, \RhSnSe{}, features an unstable $T_2^+$ phonon mode and while this leads to an increase in the unit cell, it does not alter the space group.

Motivated by the possible interplay between van Hove points and electronically driven structural transitions, we selected five of the compounds which featured van Hove points at F or L near the Fermi level while simultaneously having the flat $\Gamma$-T band further from the Fermi level. Doping or hydrostatic pressure can shift the van Hove points of these compounds to the Fermi level and in two of the cases, the materials become unstable with an imaginary L point phonon mode. \RhTlS{} becomes unstable under significant hole doping, corresponding to the amount needed to shift the electronic states at F to the Fermi level [see Figs.~\ref{fig:RhTlS}(a) and (b)]. However, this requires going beyond the region where our method of adding a background charge to the system to simulate doping is valid. In contrast, \PdSnSe{} becomes unstable for smaller values of hole doping, consistent with the regions where either the electronic states at L or at F reaches the Fermi level which, in both cases, result in an unstable L point phonon mode, as shown in Fig.~\ref{fig:PdSnSe}(c). Furthermore, the electronic states are also shifted towards the Fermi level by hydrostatic pressure, which also leads to unstable phonon modes in the system, as shown in Fig.~\ref{fig:PdSnSe}(d). To further highlight the role played by the electrons in driving these phonons unstable, we considered the impact of the electronic smearing temperature on the phonon frequencies. In the cases where the structural transitions are concomitant with van Hove points appearing at the Fermi level, i.e., for hole-doped \RhTlS{} and \PdSnSe{}, increasing the electronic smearing hardens the unstable phonon mode until it becomes stable at sufficiently large values of the smearing, as shown in Fig.~\ref{fig:ElecTemp_RhTlS} and \ref{fig:ElecTemp_PdSnSe}(a). This is in contrast to the behavior observed in \RhSnSe{} and \PdSnSe{} under pressure where the frequency of the unstable phonon is nearly unaffected by the electronic smearing [Fig.\ref{fig:ElecTemp_RhSnSe} and Fig.\ref{fig:ElecTemp_PdSnSe}(b)]. This phenomenology is to be contrasted with the \AVS{} kagome metals which feature unstable phonon modes that are also temperature-dependent~\cite{Tan2021Charge,Ratcliff2021Coherent,Christensen2021Theory,Gutierrez-Amigo2024Phonon}. However, in these cases, the parent compounds themselves are unstable~\cite{Ortiz2019New,Stahl2022Temperature-driven} and the unstable modes appear at several high-symmetry points in the Brillouin zone.

We interpret the above results to indicate that electronic correlations can drive structural instabilities in specific shandites, specifically in \RhTlS{} and \PdSnSe{}. However, moving the van Hove points at the F or L points to the Fermi level in other shandites, including \PdSnS{}, \PdPbS{}, and \RhSnSe{}, does not result in a structural transition. This implies that the electron-phonon coupling plays an important role in driving the structural transition, in line with the results of Ref.~\cite{Johannes2008Fermi}. To capture these, it may be necessary to include the effect of the motion of the apical chalcogen ions.
Nevertheless, the fact that the structural transitions in \RhTlS{} and \PdSnSe{} vanish when the electronic smearing is increased is evidence that the electronic degrees of freedom are important for the transition. This is further supported by the fact that the transitions occur once the van Hove points are moved to the Fermi level. 

Our results provide a survey of the electronic and structural properties of a large family of shandite materials. We provide indications based on first-principles calculations that it may be fertile to look for the shandite structure in Pd$_3$(In,Sn)$_2$Ch$_2$ and in Rh$_3$(In,Sn,Tl)$_2$Se$_2$, and that many of the shandites surveyed should remain free from structural instabilities up to high pressures. We hope these theoretical results will engender future experiments on the shandite compounds. As we have shown here, these materials host an interesting interplay between lattice vibrations and correlated electrons driven by the electron-phonon coupling in this kagome-layered structure.

\begin{acknowledgments}
Work at the University of Minnesota (L.B. and T.B.) were supported by the NSF CAREER grant DMR-2046020. M.H.C. is supported by ERC grant project 101164202 -- SuperSOC. Funded by the European Union. Views and opinions expressed are however those of the authors only and do not necessarily reflect those of the European Union or the European Research Council Executive Agency. Neither the European Union nor the granting authority can be held responsible for them.
\end{acknowledgments}

\section{Data Availability} 
The data that support the findings of this article are openly available at the Data Repository for University of Minnesota (Ref.~\cite{drum}).

\appendix

\section{Effect of SOC}\label{app:SOC}

\begin{figure}[!t]
    \centering
    \includegraphics[width=\linewidth]{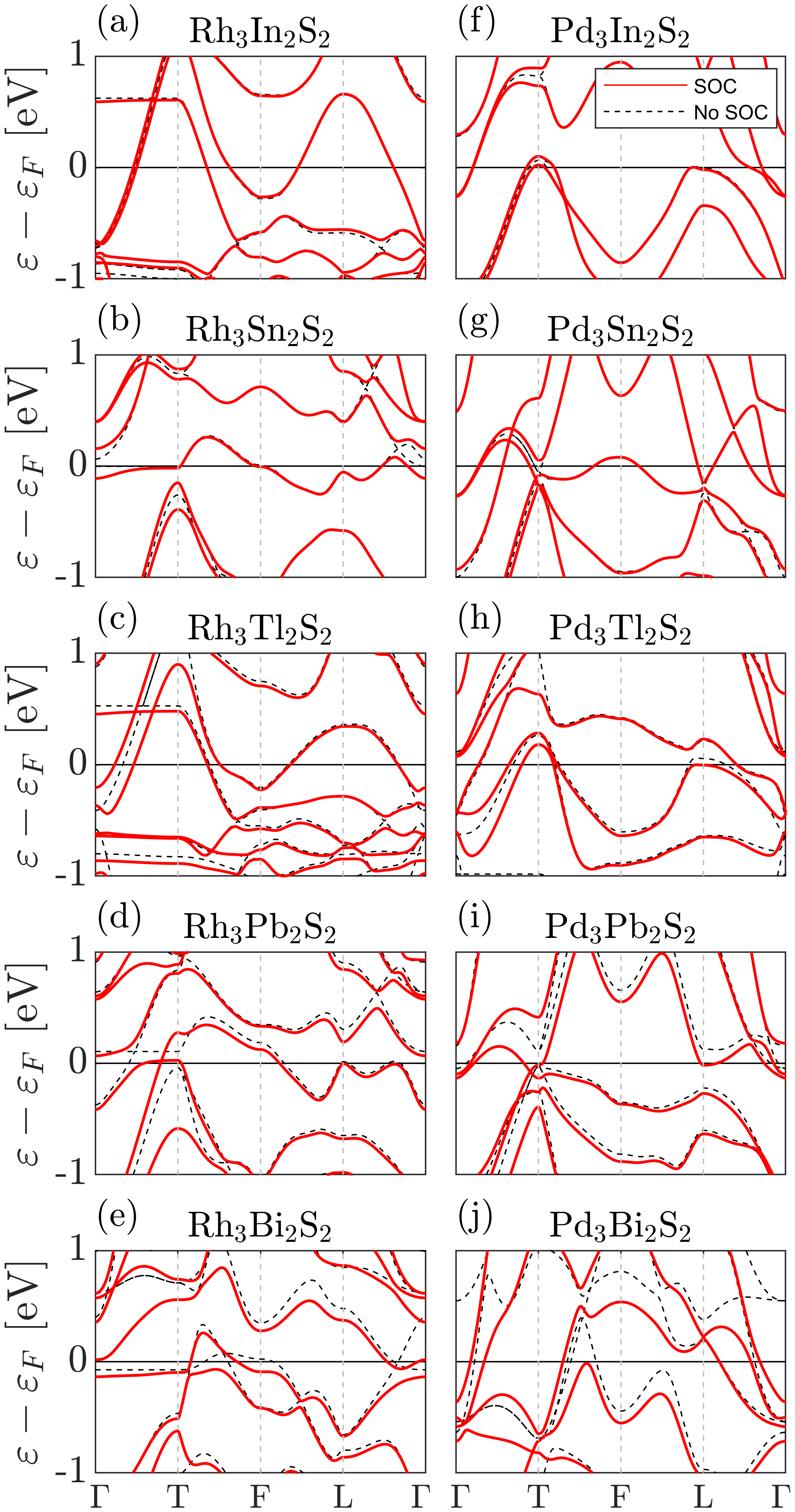}
    \caption{\label{fig:shandite_electronic_structure_soc} Effect of the SOC on the electronic band structures of shandite materials $M_3A_2$S$_2$ for (a)--(e) $M=$Rh and (f)--(j) $M=$Pd. Solid red lines denote the band structure obtained with SOC, whereas black dashed lines indicate the band structure computed without SOC.}
\end{figure}

\begin{figure}[!t]
    \centering
    \includegraphics[width=\linewidth]{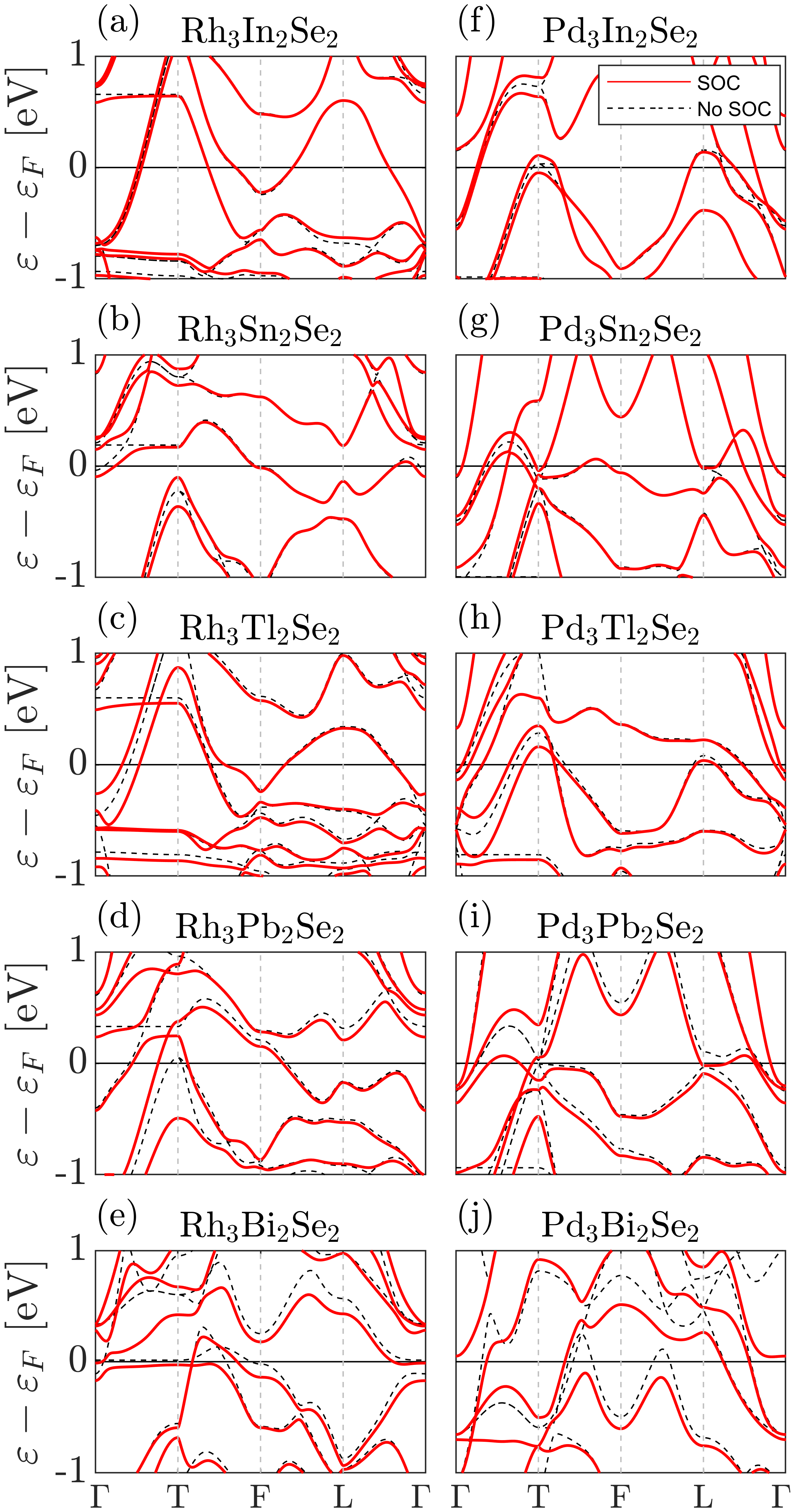}
    \caption{\label{fig:shandite_electronic_structure_soc_Se} Effect of the SOC on the electronic band structures of shandite materials $M_3A_2$Se$_2$ for (a)--(e) $M=$Rh and (f)--(j) $M=$Pd. Solid red lines denote the band structure obtained with SOC, whereas black dashed lines indicate the band structure computed without SOC.}
\end{figure}

The right-most columns of Table~\ref{tab:lattice_parameters} summarize the energy differences between the shandite and parkerite structures with and without SOC. Overall, SOC has only a minor effect on these relative energies, typically on the order of $\sim$10~meV per formula unit. Here, we examine the influence of SOC on the electronic structure of the materials in the shandite lattice structure, with particular focus on the states near the Fermi level at F and L. The corresponding band structures are shown in Fig.~\ref{fig:shandite_electronic_structure_soc} for $M_3A_2$S$_2$ and in Fig.~\ref{fig:shandite_electronic_structure_soc_Se} for $M_3A_2$Se$_2$. We find that SOC is negligible in the case of compounds with lighter $A$ elements (In, Sn, and Tl), while it leads to a few modifications for heavier $A$ elements (Pb and Bi). In most cases the states at the $F$ and $L$ points remain essentially unaffected by SOC. The most pronounced SOC effect in these materials is the splitting of the doubly degenerate $\Lambda_3$ bands along the $\Lambda$ line connecting the $\Gamma$ and $T$ points.

\bibliography{main}

\end{document}